\let\latexaddtocontents\addtocontents
\documentclass[a4paper, accepted=2023-08-23]{quantumarticle}
\let\addtocontents\latexaddtocontents
\pdfoutput=1
\usepackage{xcolor}  
\definecolor{Oorange}{HTML}{ED7D31}
\usepackage{hyperref}

\usepackage[utf8]{inputenc}

\usepackage{amsmath}      
\usepackage{amsfonts}       
\usepackage{amsthm}         
\usepackage{bbding}         
\usepackage{accents}
\usepackage{bm}             
\usepackage{graphicx}       
\usepackage{fancyvrb}       
\usepackage[nottoc]{tocbibind} 
\usepackage{dcolumn}        
\usepackage{booktabs}       
\usepackage{paralist}       
\usepackage{physics}
\usepackage{braket}         
\usepackage{bbold}          
\usepackage{xfrac}          
\usepackage{float}
\usepackage{nicefrac}
\usepackage{tikz}           
\usetikzlibrary{trees,arrows}
\usepackage{setspace}       

\usepackage[backend=bibtex, 
            isbn=false, 
            eprint=false, 
            url=false,
            sorting = none]{biblatex}
\DefineBibliographyStrings{english}{bibliography = {References}}

\title{Quantum reference frames: derivation of perspective-dependent descriptions via a perspective-neutral structure}
\author{Viktor Zelezny}
\affiliation{Institute for Quantum Optics and Quantum Information, Austrian Academy of Sciences, Boltzmanngasse 3, 1090 Vienna, Austria}

\DefineVerbatimEnvironment{code}{Verbatim}{fontsize=\small, frame=single}

\newcommand{\R}{\mathbb{R}}

\renewcommand{\H}{\mathcal{H}} 

\newcommand{\rightarro}{\quad\rightarrow\quad}

\newcommand{\ptd}[2][]{\frac{\partial #1}{\partial #2}}
\newcommand{\phiphys}{\ket{\phi}_{phys}}
\newcommand{\that}{\hat{T}}
\newcommand{\thatdagger}{\hat{T}^\dagger}

\usepackage[framemethod=tikz]{mdframed}
\usepackage{lipsum}

\begin{document}

\begingroup
\let\clearpage\relax
    \begin{abstract}
In standard quantum mechanics, reference frames are treated as abstract entities. We can think of them as idealized, infinite-mass subsystems which decouple from the rest of the system. In nature, however, all reference frames are realized through finite-mass systems that are subject to the laws of quantum mechanics and must be included in the dynamical evolution. A fundamental physical theory should take this fact seriously. In this paper, we further develop a symmetry-inspired approach to describe physics from the perspective of quantum reference frames. We find a unifying framework allowing us to systematically derive a broad class of perspective dependent descriptions and the transformations between them. Working with a translational-invariant toy model of three free particles, we discover that the introduction of relative coordinates leads to a Hamiltonian structure with two non-commuting constraints. This structure can be said to contain all observer-perspectives at once, while the redundancies prevent an immediate operational interpretation. We show that the operationally meaningful perspective dependent descriptions are given by Darboux coordinates on the constraint surface and that reference frame transformations correspond to reparametrizations of the constraint surface. We conclude by constructing a quantum perspective neutral structure, via which we can derive and change perspective dependent descriptions without referring to the classical theory. In addition to the physical findings, this work illuminates the interrelation of first and second class constrained systems and their respective quantization procedures.
\end{abstract}

    \maketitle
    \tableofcontents
    \section{Introduction}
\subsection{On quantum reference frames}
In this work, we follow the paradigm that the properties of physical systems $S$ have no absolute meaning but are only defined relatively to some other content of the universe. In theoretical classical mechanics and special relativity the role of this "other content of the universe" is usually played by abstract frames of reference. As we treat these frames as external entities, they must be non-dynamical and sufficiently decoupled from $S$. Upholding the paradigm, let us think of reference frames as abstractions of physical systems, for example ideal rigid bodies with imprinted rulers defining orientations and distances. For such reference frames to not be affected by $S$, their mass must be much larger than that of $S$, at best infinite. 

Of course, in reality, all reference frames are realised by finite-mass systems, for example some measurement devices in a laboratory. A fundamental approach should take the fact seriously that those real reference frames are subject to the laws of physics. This means that we cannot simply conjure up new frames, like we do in Galilean or Lorentz transformations. Instead, all reference frames must already be explicitly included in the dynamical evolution. Following this line of thought, reference frame transformations can be naively understood as "jumping" between distinct parts of our system.

The conviction that meaningful physical statements can only be made about the relation between things and not about the relation between things and abstract space is at the core of Mach's principle. According to one of the many circulating versions of this conjecture, Mach suggests that "Local inertial frames are affected by the cosmic motion and distribution of matter" \cite{bondi1997}. Historically, some effort was made to construct purely relational mechanics in line with Mach's principle. Two exemplary Machian frameworks can be found in \cite{barbour1977} and \cite{assis1989}.

As a second paradigm, we assume the universal validity of quantum mechanics.\footnote{Essentially, we adopt an Everettian view on quantum mechanics (see \cite{everett1957}). The question of how to interpret Everett's proposal, especially concerning the origin of probabilities, is subject to ongoing debate. The interested reader may find an overview in \cite{barrett2018}.} Now, if reference frames are realised through physical systems they must also be subject to the laws of quantum theory (see \cite{bell1990}). One can for instance imagine transforming to another laboratory being in superposition to or entangled with the present one. In accepting both paradigms, we need to face the question of how to describe physics from the perspective of \emph{quantum reference frames} (QRFs) and how to relate between such descriptions. 

QRFs have been extensively discussed in literature, with the greatest share of papers examining topics in quantum information and in finite dimensional Hilbert spaces \cite{bartlett2007,bartlett2006a,bartlett2009,boileau2008,delatorre2003,gour2008,lidar2003,palmer2014,pienaar2016,poulin2006,poulin2007}. The main focus lies here on operational aspects of QRFs, such as how to communicate quantum information without a shared frame of reference. In \cite{aharonov1967,aharonov1967a,bartlett2006} the relation of QRFs to superselection rules is discussed. Finally, the role of QRFs in infinite-dimensional Hilbert spaces and in the context of quantum foundations is investigated \cite{aharonov1984,angelo2011,angelo2012,giacomini2019,vanrietvelde2020}.

\subsection{On perspective neutral structures}
Our departing point is the paper "Quantum mechanics and the covariance of physical laws in quantum reference frames" \cite{giacomini2019}. Giacomini, Castro-Ruiz and Brukner derived the transformations between reference frames attached to quantum particles without referring to an absolute background. It is shown that there is no unique way of defining such a transformation but that its form depends on a choice of preferred coordinates. The findings shine light on what a theory of "quantum general covariance" could look like, a notion relevant in the context of quantum gravity.

The paper "A change of perspective: switching quantum reference frames via a perspective-neutral framework" \cite{vanrietvelde2020} aims to rederive the outcomes of \cite{giacomini2019} from first principles, embedding them in a structure better suited for generalization. Inspired from general relativity and quantum gravity the mathematical framework of constrained Hamiltonian systems is applied. The starting point is a Lagrangian with translational invariance. The Legendre-transformation of this Lagrangian fails to be surjective, resulting in a Hamilton theory which features a constraint of the total momentum and contains redundant variables. The resulting structure is thus interpreted as an observer-independent meta-theory which contains, so to say, all perspectives at once but is void of direct operational meaning. 

Operationally meaningful observer-dependent descriptions can then be obtained by fixing the redundancies, which classically amounts to a choice of gauge. It is shown that canonical transformations between two observer-dependent descriptions can be understood as gauge-transformations. To obtain the associated quantum picture the authors follow two distinct paths. In the \emph{reduced quantization}, the authors fix the redundancies classically before quantizing the system. This yields the quantum physics as seen from a specific observer. In the \emph{Dirac quantization} it is the other way around: One quantizes first, then fixes the redundancies by a projection on the constraint-eigenstates. This leads to a perspective-neutral quantum theory. Vanrietvelde et al. show that one can re-obtain the lot of observer-dependent descriptions by unitarily rotating the system prior to projecting such that the constraint acts only on a selected degree of freedom.

In the opinion of the present author the approach introduced in \cite{vanrietvelde2020} has one major downside: the unsymmetrical treatment of positions and momenta. While the theory features a constraint for the total momentum, there is none for the center-of-mass. A consequence of this privileged role of the momenta is that only a fraction of all perspective dependent descriptions can be obtained. Likewise, only the reference-frame transformations switching between those frames, constituting only a small subset of all possible canonical transformations, can be derived. How to embed the remaining ones is left unclear. 

\subsection{Summary of results}
\begin{figure*}[t!]
    \centering
    \includegraphics[scale=0.7]{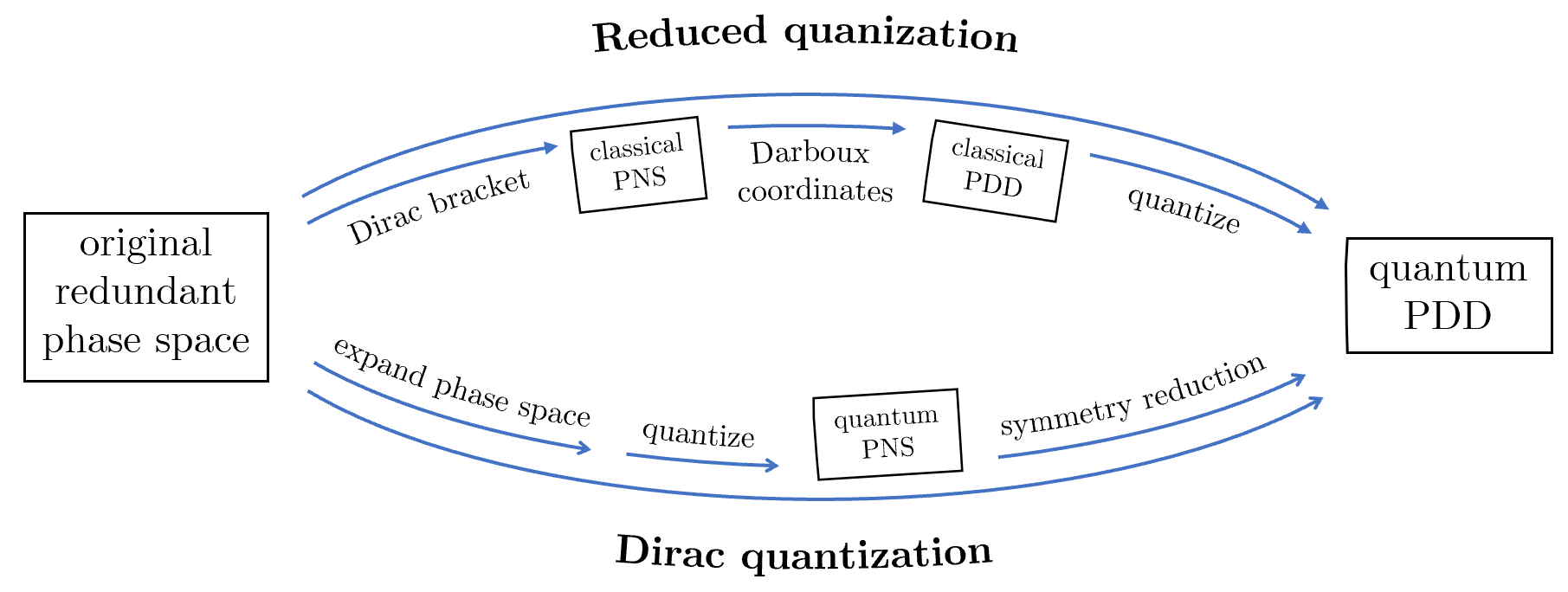}
    \caption{\small A schematic summary of the reduced and the Dirac quantization scheme. Here, PNS stands for "perspective-neutral structure" and PDD for "perspective-dependent description".}
    \label{fig: full scheme simplified}
\end{figure*}
In this article, we  aim to address the identified shortcomings of \cite{vanrietvelde2020}. We will discover that the theory created by Vanrietvelde et al. can be understood as part of a richer, more symmetrical framework where the positions and momenta are treated on equal footing. The resulting theory allows to uncover a broader set of perspective dependent frames, some of which were concealed in the original approach. Furthermore, we will systematically obtain the transformations connecting the perspective dependent frames, both in the classical and in the quantum picture. We will thus be able to derive additional transformations discussed in \cite{giacomini2019} from first principles. Besides those merits, in our extended framework, all appearing (relative) variables will be equipped with a clear physical meaning relating them to the description of an idealized outside observer. This will render the theory more accessible and further lead to a reinterpretation of some of the variables appearing in the original theory. It should be emphasized that the physics underlying the framework of the present article are equal to those of the original theory by Vanrietvelde et al., as both theories start from the same Lagrangian. We therefore expect to partly re-derive the results of \cite{vanrietvelde2020}. 

The approach proposed in this article starts with a transformation to relative coordinates. This leaves us only with gauge-independent coordinates and eliminates all references to a Newtonian background already at the Lagrangian level. The associated Hamiltonian theory is more symmetrical and features two constraints for both relative positions and momenta. Unfortunately, the constraints do not commute, thus preventing direct quantization. We tackle this difficulty following two paths. 

At first, we employ the reduced quantization scheme. Here we make use of the so-called \emph{Dirac-bracket}, which can be understood as the Poisson bracket restricted to the constraint surface. Its core advantage is that the constraints Dirac-commute, paving the way for quantization. However, this comes at a price: We will see that the Dirac bracket is not of canonical form, i.e. that phase space coordinates referring to different particles do not commute along the constraint surface in general. In the style of Vanrietvelde et al. we interpret the set of Dirac-commuting, redundant coordinates as the classical perspective-neutral structure. 

It can then be shown that the correct observer-dependent descriptions are given by those intrinsic coordinates of the constraint surface, whose Dirac bracket satisfies the canonical commutation relations. We will focus on two such descriptions: In the relative position frame two selected position coordinates match those seen from an external infinite mass-observer. In the relative momentum frame this holds for two momentum coordinates. While the former is included in \cite{vanrietvelde2020}, the latter is a new feature of the theory. The relative frames can now easily be quantized via the standard recipe. But since the quantization takes place only after having resolved the redundancies one cannot obtain a full quantum perspective neutral structure via the reduced quantization scheme. 

To fix this, we follow a second path: We further expand the phase space and  modify the constraints to make them commute. It will be shown that partially gauge fixing this extended space leaves us with a class of symplectomorphic \emph{auxiliary spaces}. One of them is identical to the perspective-neutral structure found by Vanrietvelde et al., thus proving that the original theory is fully embedded in our new framework. However, the physical interpretation of the appearing relative variables must change. There appears to be evidence that only one of the many possible gauge fixes is physically meaningful. Thus, the gauge freedom will be replaced by the freedom to choose a set of darboux coordiantes. Quantizing the extended structure then yields the sought-after quantum perspective-neutral structure.

To re-obtain the perspective-dependent quantum descriptions we carry out a quantum symmetry reduction. It consists of two unitary transformations, each followed by a projection on either constraint. We will discover that with the first unitary we decide between relative position and relative momentum description, while the second unitary determines which particle will serve as the reference frame.  Finally, we will see how the quantum perspective neutral structure can be employed to change between quantum reference frames. Figure \ref{fig: full scheme simplified} shows the connections between the various structures in a simplified scheme.

    \section{Classical Theory}
\subsection{Mathematical preliminaries: constrained  Hamiltonian systems}
\label{sec: mathematical overview}
In this section we will compactly review the mathematical formalism of constrained Hamiltonian systems, following largely \cite{henneaux1994}. The well-informed reader may skip this part. 
\subsubsection{Constraints through the Legendre transformation}

Let us write the Euler-Lagrange-equations for a given Lagrangian $L$ in their expanded form 
\begin{equation}
    \ddot q_{n'} J_{n'n} = \ptd[L]{q_n} - \dot q_{n'} \frac{\partial^2 L}{\partial q_{n'} \partial \dot q_{n}},
\end{equation}
where
\begin{equation}
    J_{n'n} =\frac{\partial^2L}{\partial \dot q_{n'} \partial \dot q_{n}}.
\end{equation}
We observe that this linear equation determines the accelerations $\ddot q_n$ uniquely only when the Jacobian is invertible, that is when the determinant 
\begin{equation}
\label{eq: Jacobian}
    det\left(\frac{\partial^2L}{\partial \dot q_{n'} \partial \dot q_{n}}\right)
\end{equation}
does not vanish. This translates directly into the Hamilton formalism since $J_{n'n}$ can be written as 
\begin{equation}
    J_{n'n} = \ptd[p_{n'}]{q_n}.
\end{equation}
We see that the non-vanishing of the determinant \ref{eq: Jacobian} is the condition for the invertibility of the velocities as function of the coordinates and momenta. 

If the rank of $J_{n'n}$ is equal to $N-M$ then the Legendre-transformation maps onto a $N-M$-dimensional submanifold of the phase-space. Only configurations contained in this subspace are physically meaningful. The constraint surface can be described by $M$ constraints
\begin{equation}
\label{eq: constraints}
    \varphi_m (q,p) \simeq 0, \quad m = 1 ... M.
\end{equation}

We have introduced the \emph{weak equality symbol} $\simeq$ to emphasize that the quantity $\varphi_m$ is numerically restricted to be zero, but does not vanish throughout phase space \cite{henneaux1994}. This means that the derivatives of $\varphi_m$ do not vanish on the constraint surface in general. Therefore, the $\varphi$'s  have in general nonzero Poisson brackets with other phase-space functions or the canonical coordinates and we must be cautious not to solve the constraints before calculating the Poisson brackets. An equation that holds everywhere in phase-space is called \emph{strong} and will be denoted by the regular equality sign "$=$".  

\begin{figure*}
    \centering
    \includegraphics[scale=0.4]{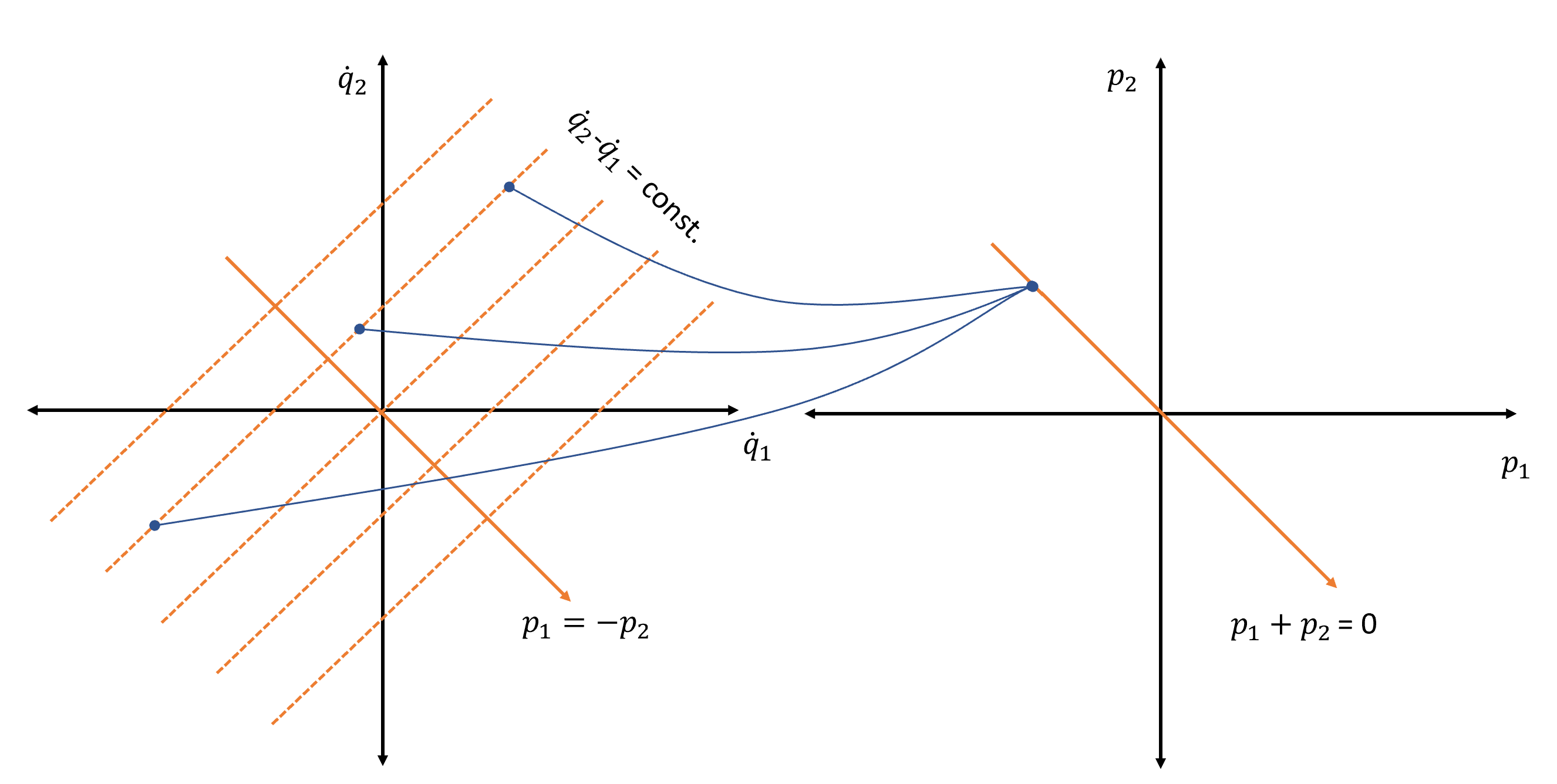}
    \caption{\small Illustration of the relation between velocities and momenta in a constrained system. The figure shows the Legendre transformation of a 2-dimensional system with the Lagrangian $L=1/2(\dot q_1 - \dot q_2)^2$. The momenta are $p_1 = \dot q_1 - \dot q_2$ and $p_2=\dot q_2 - \dot q_1$. Since $p_1 = -p_2$ the tuple $(p_1, p_2)$ describes only a 1-dimensional subspace of the $(\dot q_1, \dot q_2)$-space and all the $\dot q$'s on the line $\dot q_2 - \dot q_1 = c$ are mapped on the same point $p_2 = -p_1 = c$. We thus need additional parameters $y_m(q,\dot q)$ to render the transformation invertible.}
    \label{fig: singular lagrangian}
\end{figure*}

In the presence of constraints the variations $\delta q_n$ and $\delta p_n$ are no longer independent as we only allow for variations that are tangent to the constraint surface. We can equally state that all linear combinations of constraint variations must be weakly zero, i.e.
\begin{equation}
\label{eq: hamiltonian variation 3}
    y_m\delta \varphi _m = y_m\ptd[\varphi_m]{p_n}\delta p_n +y_m\ptd[\varphi_m]{q_n}\delta q_n \simeq 0.
\end{equation}
Comparing this equation with 
\begin{equation}
\label{eq: hamiltonian variation 2}
    \left( \ptd[H]{q_n} + \ptd[L]{q_n} \right) \delta q_n + \left( \ptd[H]{p_n} - \dot q_n\right)\delta p_n = 0. 
\end{equation}
and making use of the Euler-Lagrange equations leads to the Hamilton equation of motion for a constrained system 
\begin{align}
\begin{split}
\label{eq: eom 1}
    \dot q_n &= \ptd[H]{p_n} + y_m\ptd[\varphi_m]{p_n} = \ptd[H_T]{p_n} \\
    \dot p_n &= -\ptd[H]{q_n} - y_m\ptd[\varphi_m]{q_n} = \ptd[H_T]{q_n}.
\end{split}
\end{align}
Here we have introduced the \emph{total Hamiltonian}
\begin{equation}
    H_T = H(q, p) + y_m(q,p,t)\varphi_m(q,p)
\end{equation}
as the original Hamiltonian plus linear combinations of the constraints. 

$H_T$ is only well defined on the physical constraint surface and can be extended arbitrarily in the extended space. Choosing a specific extension corresponds to choosing the free auxiliary parameters $y_m$. As all physically sensible configurations must lie on the constraint surface, this extension does not change their respective energy values. 

With the total Hamiltonian we can write the total time derivative of any function as
\begin{align}
\begin{split}
\label{eq: eom dirac bracket}
            \dot F(q,p) &= [F,H_T] = [F,H]+[F,y_m\varphi_m]\\
               &= [F,H] + [F, y_m]\varphi_i + y_m[F,\varphi_m]\\
               &\simeq [F,H] + y_m[F,\varphi_m].
\end{split}
\end{align}

    \subsubsection{Computing the consistency conditions}
\label{sec: Computing the Consistency Conditions}
Consistency requires that the constraints be preserved in time, i.e that their total time derivative vanishes on the constraint surface. So plugging them into equation \ref{eq: eom dirac bracket} gives rise to the \emph{consistency conditions} 
\begin{align}
\begin{split}
\label{eq: consistency conditions}
    \dot\varphi_m=[\varphi_m, H] +y_{m'}[\varphi_m, \varphi_{m'}]\simeq 0,
\end{split}
\end{align}
with
\begin{equation}
     m,m' = 1,\dotso,M.
\end{equation}
These equations reduce to one of the following three types \cite{dirac1950}:

\begin{itemize}
\item[\textbf{Cycle:}] If $\varphi_m$ commutes with all $\varphi_{m'}$ but not with $H$, we obtain a relation of the form $\varphi_k(p,q) \simeq 0$. If linearly independent of the former constraints, this relation consitutes a further constraint on the system. It is evident that this new constraint must be conserved as well, leading to a new consistency condition and so forth. The cycle continues until we reach one of the stops below.

\item[\textbf{Stop 1:}] If $\varphi_m$ does not commute with at least one constraint, we do not obtain a new constraint but a relation for the $y$'s.

\item[\textbf{Stop 2:}] The equation reduces to either $0 = 0$ or $1 = 0$. The first case is automatically satisfied and leads to no further constraint, while the second one points to an inconsistency in the equations of motions. Such theories start from self-contradictory Lagrangians and are of no interest.
\end{itemize}

Having exhausted all equations we are left with a complete set of $J$ constraints. We can add them to the total Hamiltonian and let the consistency conditions run now over all constraints
\begin{equation}
\label{eq: consistency conditions 2}
    [\varphi_j, H] +y_{j'}[\varphi_j, \varphi_{j'}]\simeq 0, \quad j,j' = 1,\dotso,J.
\end{equation}
This set of linear equation is over-complete, since a part of it has already been solved to derive the secondary constraints. We see that the solutions depend crucially on the matrix of commutators
\begin{equation}
\label{eq: matrix of commutators}
    \Delta_{jj'} = [\varphi_j,\varphi_{j'}].
\end{equation}
If it is invertible then we can find unique solutions for all $y_j$ and thus the Hamiltonian and the time evolution of the system is unique too. If, however, $\Delta_{jj'}$ has a vanishing determinant, then there are some zero eigenvectors and not all $y_m$ are uniquely determined. 

Let us quickly introduce some new terminology. In \cite{dirac2001}, Dirac referred to any function on phase-space as \emph{first class}, whose Poisson bracket with every constraint is weakly equal zero
\begin{equation}
    [F,\varphi_j] \simeq 0 \quad j = 1 ...J.
\end{equation}
The Hamiltonian, for example, is first class by design. A function which has at least one non-vanishing Poisson bracket with any constraint is called \emph{second class}.

We now perform linear transformations $\varphi_j \rightarrow a_{jj'}\varphi_{j'}$ on the constraints. 
Our aim is to split the such transformed constraints into a subset $\gamma_\alpha$ of \emph{first class constraints} and a subset $\Phi_\alpha$ of \emph{second class constraints}. As combinations of weakly vanishing quantities also vanish weakly, the linear transformation does not change the theory. After having successfully carried out the transformation we can write the matrix of commutators in the form
 \begin{gather}
 \label{eq: matrix of commutators 2}
    \Delta =
    \begin{pmatrix}
    [\gamma_a,\gamma_b]&[\Phi_\alpha,\gamma_b]\\[4pt]
    [\gamma_a,\Phi_\beta]&[\Phi_\alpha,\Phi_\beta]
    \end{pmatrix}
    =
    \begin{pmatrix}
    0&0\\
    0&C_{\alpha\beta}
    \end{pmatrix}
\end{gather}

The first class constraints $\gamma_a$ do not impose any condition on their respective $y_a$, which means that $y_a$ are arbitrary functions of the phase-space coordinates and time. The second class constraints $\Phi_\alpha$, on the contrary, have well defined $y_\alpha$, which we can compute by means of the equations \ref{eq: consistency conditions 2}. Finally we can write our new total Hamiltonian as the sum of the original Hamiltonian plus linear combinations of all constraints
\begin{align}
\begin{split}
    H_T &= H + y_\alpha\Phi_\alpha + y_a(p,q,t)\gamma_a \\
        &= H'+ y_a(p,q,t)\gamma_a,
\end{split}
\end{align}
where
\begin{equation}
     H' = H + y_\alpha\Phi_\alpha.
\end{equation}

\subsubsection{Gauge transformations by first-class constraints}
\label{sec: Gauge Transformations by first-class constraints}
Note that the above total Hamiltonian $H_T$ decomposes into two parts. While $H'$, containing the original Hamiltonian plus the second-class constraints, generates a deterministic evolution, the contribution from $y_a(p,q,t)\gamma_a$ is completely arbitrary. 

Let us take e.g. $f(q,p) = q$ and compute its value after an infinitesimal time span for distinct values of the $y_a$. We will obtain two different outcomes, their difference being
\begin{equation}
    \Delta q(t_0 + \delta t) =  \delta t(y_a - y'_a)[q, \gamma_a].
\end{equation}
As the time-dependence of the functions $y_a$ is not accessible, we have no means of saying anything about the evolution of $q$. This is of course incompatible with the deterministic character of classical mechanics. If we want unique solutions, it must be that operationally we cannot distinguish between two configurations $(q_i, p_i)$ and $(q'_i, p'_i)$ connected by a so-called \emph{gauge transformation}
\begin{align}
\begin{split}
\label{eq: gauge transformation}
            q_i \rightarrow q'_i &= q_i + y_a [q_i, \gamma_a]\\
            p_i \rightarrow p'_i &= p_i + y_a [p_i, \gamma_a].\\
\end{split}
\end{align}
This means that even though a physical state is fully determined by the variables $(q,p)$, the converse is not true: A state does not uniquely determine a point in phase-space of $q$'s and $p$'s but a whole set of points. From equation \ref{eq: gauge transformation} we learn that the gauge transformations connecting these points are generated by the first class constraints, which is in accordance with Noether's theorem \cite{kosmann2011noether}. 

Furthermore, not all functions on phase-space correspond to physically measurable quantities. It is only the gauge invariant functions, that are those whose Poisson bracket  with every first-class constraint vanishes weakly, which can be regarded as physically meaningful observables. 

Second class constraints do not act as gauge generators, as they would map states out of the constraint surface. This can be seen by taking any second class constraint $\Phi_\alpha$ and checking the transformation induced by any other second class constraint $\Phi_\beta$

\begin{equation}
    \Phi_\alpha \rightarrow \Phi_\alpha' = \Phi_\alpha + \epsilon [\Phi_\alpha, \Phi_\beta]  \not\simeq 0.
\end{equation}

\subsection{Transformation to relative coordinates}
In this section, we will work out the central structure of our paper, the second-class constrained Hamiltonian system. The departing point is a toy-model Lagrangian of three free particles in one dimension, reading
\begin{equation}
\label{eq: Lagrangian}
    L(x,\dot{x}) = \frac{1}{2}\sum_{i=1}^3 m_i\dot{x}_i^2-\frac{M}{2}\dot x^2_{cm},
\end{equation}
where $$x_{cm} = (\sum_{i=1}^3 m_i x_i)/M$$ is the center-of-mass-position and $$M= \sum_{i=1}^3 m_i$$ the total mass of the system. By subtracting the center-of-mass kinetic energy we have rendered the Lagrangian invariant under local translations
$$
(q_i,\dot q_i)\rightarrow (q_i+f(t),\dot q_i+\dot f(t))
$$
where $f(t)$ is an arbitrary function of time. Because of this gauge symmetry the physical description depends only on relative degrees of freedom but not on the choice of external reference frame \cite{vanrietvelde2020}. 

\begin{figure*}
    \centering
    \includegraphics[scale=0.6]{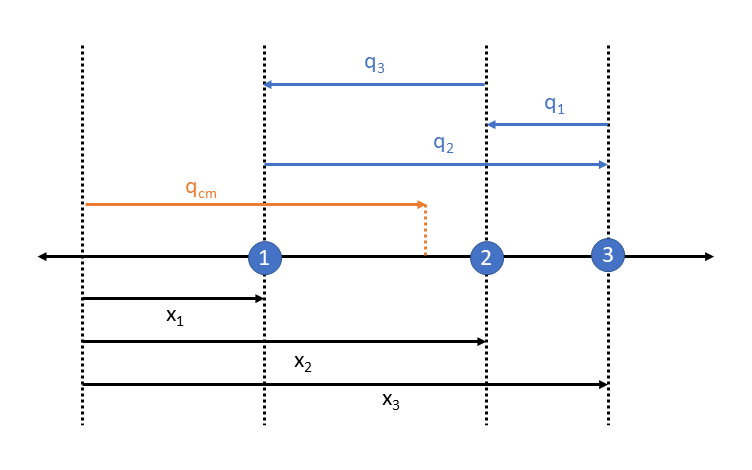}
    \caption{\small The figure shows the transformation \ref{eq: coordinate transform} schematically. We notice that after the transformation $q_{cm}$ remains the only coordinate with respect to an external observer and that the $q_i$ are redundant.}
    \label{fig:my_label}
\end{figure*}

The starting idea of this paper is to make the reference-frame independence explicit by going over to the coordinates
\begin{align}
\label{eq: coordinate transform}
\begin{split}
    x_1-x_2 &=q_3\\
    x_3-x_1 &= q_2\\
    x_2-x_3 &= q_1\\
    \frac {1}{M}\sum_{i=1}^3 m_i x_i&= q_{cm}.
\end{split}
\end{align}
The three relative coordinates $q_i$ are invariant under local spatial translations, while the center-of-mass coordinate remains gauge dependent and still refers to the Newtonian background. 

In geometrical terms the transformation \ref{eq: coordinate transform} corresponds to a mapping from $\mathbb{R}^3$ into $\R^4$, reaching only points on the 3-dimensional hypersurface
\begin{equation}
\label{eq: constraint Q}
    q_1+q_2+q_3 \equiv Q = 0.
\end{equation}
It follows that the $q_i$ are redundantly defined; we can always express one of the relative distances by means of the other two.\footnote{In classical mechanics constraints are often introduced to approximate very strong and short-ranged forces (see for example \cite{jensen1971}). Not so in this case, where the constraint is truly a logical necessity and any deviation from the constraint surface is prohibited by the definition of the coordinates.} 

The Lagrangian \ref{eq: Lagrangian} expressed in the new coordinates reads
\begin{equation}
\label{eq: Lagrangian new}
L'(q, \dot q) = \frac{1}{2}\sum_{i=1}^3 \mu_i\dot q_i^2 + q_4(q_1+q_2+q_3)
\end{equation}
where we have introduced the quantities
\begin{equation}
    \mu_i = \frac{m_jm_k}{M}, \quad i\neq j\neq k,
\end{equation}
which are called reduced masses and added the constraint $Q$ together with a Lagrange multiplier $q_4$. The calculation is given in appendix \ref{apsec:rewriting the lagrangian}.

Given only the Lagrangian \ref{eq: Lagrangian new}, the role of $q_4$ as a Lagrange multiplier is not apparent. It follows that $q_4 = q_4(t)$ can be treated as just another dynamical variable and that it enters the formalism on equal footing with the first three $q_i$  \cite{prokhorov2011a}. We observe that the Lagrangian, as expected, does not depend on the gauge-dependent center-of-position $q_{cm}$. This degree of freedom is of no physical relevance and will be discarded.

Let us now apply the methods established throughout section \ref{sec: mathematical overview} to obtain the Hamilton theory. We start by computing the conjugate momenta
\begin{align}
\begin{split}
    p_i &= \ptd[L]{\dot q_i}=\mu_i\dot q_i \quad i = 1,2,3\\
    p_4 & \simeq 0.  
\end{split}
\end{align}
There is one constraint $p_4\simeq 0$, which is added with yet another Lagrange multiplier $y_i$ to the standard Hamiltonian $H'$ yielding
\begin{align}
\begin{split}
\label{eq: full Hamiltonian}
    H &\simeq H'(q, p) + y_i \Phi_i \\
       &\simeq \sum_i p_i \dot q_i(q,p) - L(q,\dot q_i(q,p)) + y_i\Phi_i\\
       &\simeq \frac{1}{2}\sum_{i=1}^3 \left(\frac{p_i^2}{\mu_i} - q_4q_i\right) + yp_4.
\end{split}
\end{align}

Starting from Hamiltonian \ref{eq: full Hamiltonian} we need to work through the consistency scheme discussed in section \ref{sec: Computing the Consistency Conditions}. The calculation, done in detail in appendix \ref{apsec: derivations of consistency conditions}, yields the Hamiltonian
\begin{equation}
\label{eq: final Hamiltonian}
    H =\frac{1}{2}\sum_i \frac{p_i^2}{\mu_i}.
\end{equation}
together with 2 effective constraints 
\begin{equation}
\label{eq: 2 constraints}
    \Phi_1 = \sum_iq_i, \quad \Phi_2 = \sum_i \frac{p_i}{\mu_i}. 
\end{equation}
In the \emph{original phase-space} $\mathcal{P}_{orig} \sim \R^6$, coordinatized by the $q_i$ and $p_j$, they define a 4-dimensional constraint surface $\Phi_1 \simeq \Phi_2 \simeq 0$, on which all physically sensible configurations must lie. Clearly, they are second-class as 
\begin{equation}
    [\Phi_1, \Phi_2] = \sum_i \frac{1}{\mu_i} .
\end{equation}

It can be shown that the structure \ref{eq: final Hamiltonian} \& \ref{eq: 2 constraints} is symplectomorphic to the perspective neutral structure proposed in \cite{vanrietvelde2020}, i.e. to 
\begin{equation} 
\label{eq: perspective neutral structure vanrietvelde}
    H =\frac{1}{2}\sum_i \frac{\tilde p_i^2}{m_i}, \quad \Phi_1 = \sum_i \tilde p_i,
\end{equation}
if the latter is gauge fixed by 
\begin{equation}
\label{eq: gauge vanrietvelde}
    \tilde\Phi_2 = \sum_i \mu_i x_i.\footnotemark
\end{equation}

\addtocounter{footnote}{-1}
\footnotetext{The canonical transformation is induced by
\begin{equation*}
\label{eq: transformation vanrietvelde viktor}
    x_i = \frac{\mu_i}{\sqrt{\kappa}}q_i, \quad \tilde p_i = \frac{\sqrt{\kappa}}{\mu_i}p_i,
\end{equation*}
where
\begin{equation*}
\label{eq: definition kappa}
    \kappa = \frac{m_1m_2m_3}{m_1+m_2+m_3}.
\end{equation*}}
In the following sections, we will see that this choice of gauge is critical for the construction of our version of the perspective neutral structure which eventually leads to the derivation of further perspective-dependent descriptions. In fact, it is the critical difference between our work and \cite{vanrietvelde2020}, where the system was gauge-fixed by $q_i=0$. There, the gauge fix \ref{eq: gauge vanrietvelde} must seem arbitrary as there is no reason to assume a preferred role for this specific observer. In our scheme, however, it becomes clear, that it is exactly \ref{eq: gauge vanrietvelde} which gives meaning to the variables as relative coordinates. In other words: if one starts with relative variables, \ref{eq: gauge vanrietvelde} arises as necessary consequence of the over-parametrization of the configuration space.

    \section{Reduced quantization of second class constraints}
\label{chap: Quantization of second class constraints}

\subsection{Mathematical preliminaries: Dirac bracket and Darboux coordinates}
\label{sec: The Dirac Bracket}
If there are no constraints present and the phase-space is linear there is a standard quantization recipe (see e.g. Dirac \cite{dirac1981principles}, p. 84ff). It tells us to assign to each function an operator acting on some Hilbert space $\H$ such that the commutator of quantum operators  equals $i\hbar$ times the operator of the Poisson bracket

\begin{equation}
\label{eq: correspondace rule}
    [\hat F, \hat G] = i\hbar \widehat{[F,G]}.
\end{equation}

In the presence of second-class constraints the quantization procedure is more involved. Classically, the constraints restrict the system to a submanifold of the original phase space. We would like to mirror this behaviour in the quantum theory by imposing the restrictions
\begin{equation}
\label{eq: constraint restrictions}
    \hat \Phi_\alpha\ket{\phi}_{phys} = 0 \quad \forall \; \alpha
\end{equation}
on the state vectors $\phiphys$ which span the physical Hilbert space $\H _{phys}$. From the consecutive application of constraints  $\hat \Phi_\alpha  \hat \Phi_\beta\ket{\phi}_{phys} = 0$ and $\hat\Phi_\beta\hat\Phi_\alpha\ket{\phi}_{phys} = 0$ it follows readily that 
\begin{equation}
\label{eq: second class commutator}
    [\hat \Phi_\alpha,\hat \Phi_\beta]\phiphys = 0.
\end{equation}
The commutator of second class constraints however is not a constraint itself so that equation \ref{eq: second class commutator} is only satisfied when $\phiphys = 0$. But then $\H_{phys} = \{0\}$ and the theory is trivial \cite{kempf2001}. 

The difficulties originate already in the classical theory: While the constraints vanish on the constraint surface, their first derivatives in general do not, leaving the Poisson bracket ambiguous. This means, for a function $f$ on the phase space and a constraint $\Phi \simeq 0$ we have $f+\Phi \simeq f$, but in general
\begin{equation}
    [f,g] \not \simeq [f+\Phi, g].
\end{equation}

A solution to both the problems in classical and quantum theory was developed by P.A.M. Dirac in 1950 (see \cite{dirac1950} and for a detailed derivation \cite{dirac2001}). He generalized the original Poisson bracket to the \emph{Dirac bracket}, given by 
\begin{equation}
\label{eq: Dirac bracket}
    \{F,G\}=[F,G] -[F,\Phi_\alpha]C^{\alpha\beta}[\Phi_\beta,G]
\end{equation}
where $C^{\alpha\beta}$ is the inverse of the second-class bracket-matrix $C_{\alpha\beta}$ appearing in \ref{eq: matrix of commutators 2}. The core advantage is that the Dirac bracket of an arbitrary function with any second class constraint vanishes:
\begin{equation}
\label{eq: DB second class constraint}
\{\Phi_\alpha, F\} = 0.
\end{equation}

From a geometrical viewpoint, the Dirac bracket is related to the pull-back of the symplectic two-form onto the constraints surface and picks up only those variations tangent to it \cite{henneaux1994}. As a consequence, there are in general always contributions from non-Poisson-commuting observables. This entails that Poisson-commuting observables are in general not Dirac-commuting. We will see this concretely in the next section. 

By virtue of \ref{eq: DB second class constraint}, the Dirac bracket, unlike the Poisson bracket, is unambiguous, i.e. 
$$
f'\simeq f \rightarrow \{f',\cdot\} \simeq \{f,\cdot\}, 
$$  
such that we can set the constraints to zero before computing the bracket. To quantize a second-class system, we simply replace the Poisson bracket by the Dirac Bracket in the correspondence rule
\begin{equation}
\label{eq: dirac correspondence rule}
    [\hat F, \hat G] = i\hbar \widehat{\{F,G\}}.
\end{equation}
The constraint operators can now be given any c-values, in particular $\Phi_\alpha = 0$, enforcing the restriction \ref{eq: constraint restrictions}. When we have found a representation of \ref{eq: dirac correspondence rule} we have found the quantum theory.

\begin{itemize}
    \item[\textbf{Caveat:}] This is a highly nontrivial problem which has no general solution. It can be solved only in special instances, for example when the Dirac bracket amounts to c-numbers. Luckily, this is exactly the case for our system at hand.
\end{itemize}

There is yet an alternative to trying to solve \ref{eq: dirac correspondence rule} directly. With the Dirac bracket, the second-class constraints can be treated as strong identities expressing some canonical coordinates in terms of others. This allows us to go over to \emph{intrinsic} (local) coordinates $u_i(q,p),\pi_i(q,p), i = 1,\dots, N-M$ spanning the constraint surface. Expressed in the new coordinates the constraints $\Phi_\alpha(q_n(u_i,\pi_i),p_n(u_i,\pi_i))$ vanish identically \cite{klauder1998}. Of special interest are those intrinsic coordinates which Dirac-commute according to the canonical commutation relations
\begin{align}
        \{u_i(q,p),\pi_j(q,p)\} &= \delta_{ij}\\
        \{u_i(q,p),u_j(q,p)\}&= \{\pi_i(q,p),\pi_j(q,p)\} =0.
\end{align}
They are referred to as \emph{Darboux coordinates} \cite{klauder1998}. When they are found, we have obtained a \emph{reduced} theory, with no constraints present and no redundancies left. The Darboux coordinates are nicely canonical so that from this point on we can follow the standard quantization procedure to obtain the quantum theory. 

\begin{itemize}
    \item[\textbf{Caveat:}] Darboux coordinates are only defined up to a canonical transformation, but canonical quantization and canonical transformations are generally non-commutative operations. They do commute only for linear and point transformations (in both $q$ and $p$) \cite{castellani1979}.
\end{itemize}

Last but not least it is worth noticing that the Dirac bracket with any first class constraint is weakly equal to the Poisson bracket. Thus the equations of motions remain invariant under the replacement
\begin{equation}
    \dot F = \{F,H\} \simeq [F,H].
\end{equation}

\subsection{The classical perspective-neutral structure}
We will now apply the reduced quantization scheme to our theory at hand. At first, let us compute the Dirac-bracket. The bracket-matrix \ref{eq: matrix of commutators} reads
\begin{align}
    C = 
    \begin{pmatrix}
     0&\mu\\
     -\mu&0\\
    \end{pmatrix}
\end{align}
where we have introduced 
\begin{equation}
\label{eq: sum of virtual masses}
    \mu = \sum_i \frac{1}{\mu_i}.
\end{equation}
Since either $C^{-1}$ or the Poisson bracket of coordinates and constraints vanish, most of the terms in \ref{eq: Dirac bracket} drop out and we obtain
\begin{align}
\label{eq: computed dirac bracket}
    \begin{split}
        \{q_i,q_j\} &= 0 \\
        \{p_i,p_j\} &= 0 \\
        \{q_i,p_j\} &= \delta_{ij} -\frac{1}{\mu \mu_i}.
    \end{split}
\end{align}

In the style of Vanrietvelde et al. I now propose to interpret the Dirac bracket \ref{eq: computed dirac bracket} along with the constraints \ref{eq: 2 constraints} and the Hamiltonian \ref{eq: final Hamiltonian} as the \emph{classical perspective-neutral structure}, living in the 6-dimensional phase space $\mathcal{P}_{orig}.$ As all relative variables are treated on equal footing it contains all perspectives at once while the inherent redundancies prevent an immediate operational interpretation.

Note that position and momentum observables of different particles do not Dirac-commute. This follows from the fact, that the $q_i$'s and $p_i$'s form a redundant over-parametrization of the phase space. As we restrict to the constraint surface, a variation of one of the $q_i$'s ($p_i$'s, respectively) necessarily entails some variation of the other relative position or momentum coordinates.

This has the seemingly counter-intuitive consequence that the Dirac-bracket of two $q$'s and two $p$'s, naively chosen as intrinsic coordinates, does not fulfill the canonical commutation relations, i.e., that this choice of intrinsic coordinates is non-Darboux. In the quantum picture, the impossibility to find simultaneous eigenvectors of the $\hat q$'s and $\hat p$'s even for distinct particles follows. Therefore, from the perspective of one particle the Hilbert space of the two other particles can not be partitioned as $H_{i,j} = H_{i} \otimes H_{j}$ and the standard quantization recipe is bound to fail.

Thinking in physical terms, the commutation relations \ref{eq: computed dirac bracket} mean that $q_i$ and $p_j$  cannot be measured simultaneously and that there exist uncertainty relations between those variables. But why should a position measurement of one particle affect the momentum of another unrelated particle? The solution to this alleged paradox lies again in the finite mass $m$ of the system serving as reference frame and has already been presented in \cite{aharonov1984, angelo2011, angelo2012}. It follows from the uncertainty principle that if one wants to measure the position of a particle with an accuracy $\Delta x$ a finite amount of momentum 
\begin{equation}
    \Delta p > \frac{\hbar}{\Delta x}
\end{equation}
has to be exchanged. The system serving as reference frame is therefore boosted by an uncertain amount $\frac{\Delta p}{m}$. In classical mechanics one can always keep the amount of exchanged momentum arbitrary small, such that $\Delta p << m$. In the quantum regime, however, this approximation is no longer valid as masses are small and $\Delta p$ increases with measurement accuracy. Here, one can always envisage a measurement where $\Delta p \sim m$, resulting in a non negligible kickback of the reference system \cite{aharonov1984}.

\begin{figure*}
    \centering
    \includegraphics[scale=1]{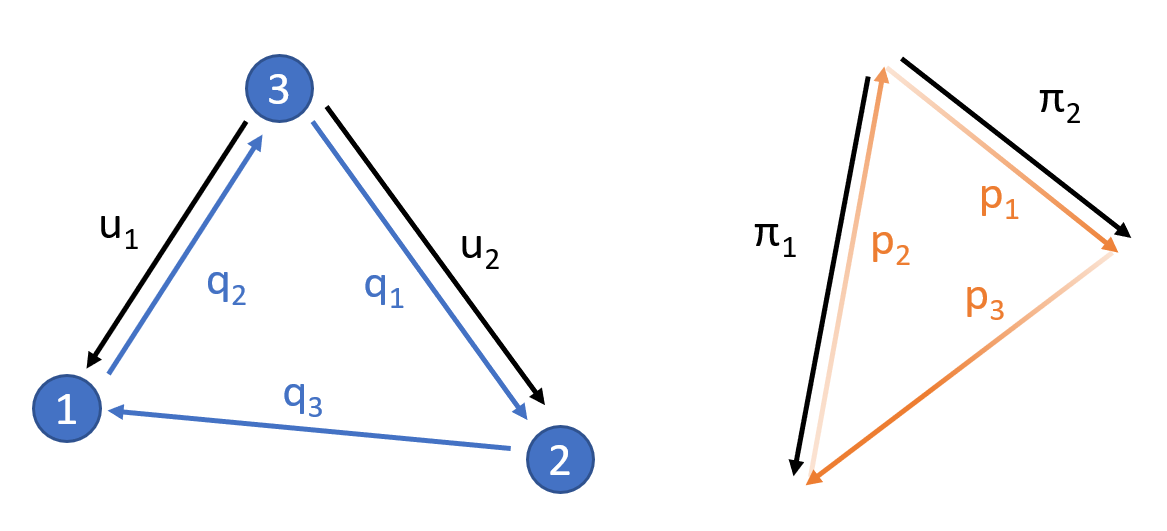}
    \caption{\small The figure shows a naive choice of internal coordinates $(u_i, \pi_j)$ that do not Dirac-commute according to the canonical commutation relations. The position vectors $q_i$ on the left hand side live in the configuration space. Displayed on the right hand side are the momentum vectors $p_j$, living in momentum space.}
    \label{fig: naive relative coordinates}
\end{figure*}

Note that there is no configuration of the masses which reproduces the standard bracket structure for all three coordinate pairs. If, however, one of the particles masses is much larger than the others, this particle can serve as a classical reference frame.  For example, for $m_1 >> m_2 \sim m_3$ the mass term $\frac{m_i}{M}=\frac{1}{\mu\mu_i}$ converges to zero and we re-obtain the canonical commutation relations. Further, $\mu_1 \rightarrow 0$ and $\mu_2\rightarrow m_3,\mu_3\rightarrow m_2$  such that the perspective-neutral Hamiltonian \ref{eq: final Hamiltonian} converges to the standard Hamiltonian for a two-particle system. This is an important result. It shows that the standard way of describing physics from an infinite-mass frame is fully embedded in our newly developed framework.

\subsection{Perspective-dependent frames}
In the section above, we saw that not all sets of intrinsic coordinates are physically useful. Thus, if we want observer-dependent descriptions which can be canonically quantized for arbitrary mass configurations we must search for Darboux coordinates. To keep it simple and not encounter any mathematical pitfalls we restrict to those which are linear in the original positions and momenta and do not mix them. Their most general form is given by
\begin{align}
\begin{split}
    \begin{pmatrix}
    q_1\\q_2\\q_3
    \end{pmatrix}
    =
    \begin{pmatrix}
    m_{1A}u_A+m_{1B}u_B\\ m_{2A}u_A+m_{2B}u_B\\ m_{3A}u_A+m_{3B}u_B
    \end{pmatrix}
    \\
    \begin{pmatrix}
    p_1\\p_2\\p_3
    \end{pmatrix}
    =
    \begin{pmatrix}
    n_{1A}\pi_A+n_{1B}\pi_B\\ n_{2A}\pi_A+n_{2B}\pi_B\\ n_{3A}\pi_A+n_{3B}\pi_B
    \end{pmatrix}.
    \end{split}
\end{align}
The 12 coefficients must satisfy 4 constraints 
\begin{equation}
\label{eq: constraint conditions}
\begin{aligned}
    \sum_i m_{iA} &= 0&\qquad \sum_i \frac{n_{iA}}{\mu_i} &= 0\\
    \sum_i m_{iB} &= 0&\qquad \sum_i \frac{n_{iB}}{\mu_i} &= 0
    \end{aligned}
\end{equation}
as well as 9 Darboux conditions
\begin{equation}
\label{eq: darboux conditions}
    m_{iA}n_{jA}+ m_{iB}n_{jB} = \delta_{ij}-\frac{1}{\mu\mu_i}
\end{equation}
with 
\begin{equation}
    i,j =1,2,3.
\end{equation}
The Darboux conditions enforce the canonical commutation relations between the intrinsic coordinates. Of these 13 constraints only 8 are linearly independent, leaving us with 4 free coefficients.

Still, we are left with a plethora of possible descriptions. How do we choose among them? I propose to coordinatize the constraint surface by two of the $q_i$ or $p_j$ and then to complete the remaining coordinates according to \ref{eq: constraint conditions} - \ref{eq: darboux conditions}. Following this line of thought, I want to lay emphasis on two internal descriptions. For $m_1\rightarrow\infty$ both of them will converge to the "classical" description from an external infinite-mass frame. 

\subsubsection{The relative position frame}
Let us choose $m_{2A} = - m_{3B} = 1$ and $m_{2B} = m_{3A} = 0$ and set the placeholders to $A=3, B=2$, such that the position coordinates are named according to figure \ref{fig: naive relative coordinates}. By applying the constraints and working through the Darboux conditions we arrive at
\begin{align}
\begin{split}
\label{eq: darboux relative position coordinates}
    \begin{pmatrix}
    q_1\\q_2\\q_3
    \end{pmatrix}
    &=
    \begin{pmatrix}
    u_2-u_3\\u_3\\-u_2
    \end{pmatrix}
    \\ 
    \begin{pmatrix}
    p_1\\[4pt]
    p_2\\[4pt]
    p_3
    \end{pmatrix}
   &= \frac{1}{\mu}
    \begin{pmatrix}
    (\frac{\pi_2}{\mu_3}-\frac{\pi_3}{\mu_2}) \\[4pt]
    \frac{1}{\mu_3}\pi_2+(\mu - \frac{1}{\mu_2})\pi_3\\[4pt]
    - (\mu - \frac{1}{\mu_3})\pi_2 -\frac{1}{\mu_2}\pi_3 
    \end{pmatrix}.
\end{split}
\end{align}
In this description we have a direct expression for the distances between particle 1 \& 3 and 1 \& 2, respectively, while the distance between 2 \& 3 can only be indirectly computed through $u_2-u_3$. It seems therefore reasonable to interpret \ref{eq: darboux relative position coordinates} as the physics seen from the perspective of particle 1. The internal position observables are in accordance with those seen from an external infinite-mass frame, while the internal momentum observables are linear combinations thereof. By such reasoning we name this set of variables the \emph{relative position frame of particle 1}. 

Plugging \ref{eq: darboux relative position coordinates} into the perspective-neutral Hamiltonian \ref{eq: final Hamiltonian} yields
\begin{multline}
\label{eq: relative position hamiltonian}
     H_{q1} = \frac{1}{2\mu}\Big(
    (\frac{\mu}{\mu_3}-\frac{1}{\mu_3^2})\pi_2^2 +
    (\frac{\mu}{\mu_2}-\frac{1}{\mu_2^2})\pi_3^2 + 2\frac{\pi_2\pi_3}{\mu_2\mu_3}\Big)  
\end{multline}
which looks tidier when expressed in the original masses 
\begin{equation}
    H_{q1} = \frac{1}{2}\left(\frac{m_1+m_2}{m_1m_2}\pi_2^2 + \frac{m_1+m_3}{m_1m_3}\pi_3^2 + 2\frac{\pi_2\pi_3}{m_1}\right).
\end{equation}
The index q next to $H$ indicates that we are we are in the relative position representation and the 1 that we are using particle 1 as reference frame. Observe that apart from the potential energy, which we have assumed to be zero, our outcome is identical to the findings of \cite{vanrietvelde2020}. The modification to other particles follows readily.

\subsubsection{The relative momentum frame}
Let us now choose $n_{2A} = - n_{3B} = 1$ and $n_{2B} = n_{3A} = 0$ and set the placeholders again to $A=3, B=2$, such that the momentum coordinates are named according to the right side of figure \ref{fig: naive relative coordinates}. In this setup we coordinatize the constraint surface by
\begin{align}
\begin{split}
\label{eq: darboux relative momentum coordinates}
    \begin{pmatrix}
    q_1\\q_2\\q_3
    \end{pmatrix}
    &=\frac{1}{\mu}
    \begin{pmatrix}
    \frac{1}{\mu_1}(v_3-v_2)\\ 
    \frac{v_2}{\mu_2} + (\mu - \frac{1}{\mu_2})v_3\\
     -(\mu - \frac{1}{\mu_3})v_2 - \frac{v_3}{\mu_3}
    \end{pmatrix}
    \\
    \begin{pmatrix}
    p_1\\p_2\\p_3
    \end{pmatrix}
    &=
    \begin{pmatrix}
    -\mu_1(\frac{\rho_3}{\mu_2}-\frac{\rho_2}{\mu_3})\\ \rho_3\\ -\rho_2.
    \end{pmatrix}
\end{split}
\end{align}
Observe that now only the momenta of particle 2 \& 3 correspond to the external view, while the remaining momentum and all positions are expressed through linear combinations of the $v$'s and $\rho$'s. We will call this set of internal observables the \emph{relative momentum frame of particle 1}. In this frame the Hamiltonian reads
\begin{multline}
\label{eq: relative momenta hamiltonian}
    H_{p1} = \frac{1}{2}\Big(\frac{\mu_1+\mu_2}{\mu_2^2}\rho_3^2 + \frac{\mu_1+\mu_3}{\mu_3^2}\rho_2^2-\frac{2\mu_1}{\mu_2\mu_3}\rho_2\rho_3\Big).
\end{multline}

The relative momentum frame cannot be derived directly via the framework of \cite{vanrietvelde2020}. This shows that while encompassing its precursor's finding, the approach proposed in this article shines further light on the theory of QRFs.

\subsection{Changing quantum reference frames}
Our next objective is to find the connection between canonical QRF-transformations and the perspective neutral structure. In contrast to the work of Vanrietvelde et al., in our framework, reference frame transformations cannot amount to gauge transformations, as there is no gauge-freedom in purely second-class systems. In exchange, we have the freedom to choose any Darboux-parametrization of the constraint surface, each of which is corresponding to a perspective dependent description. As all sets of Darboux coordinates are connected via a canonical transformation, we can conclude: To canonically transform between two perspective dependent descriptions means to Darboux-reparametrize the constraint surface. 

Finding the (quantum) canonical transformation connecting two descriptions is a three-step-process:
\begin{enumerate}
    \item We write the original, relative phase space variables $(q,p)$ in terms of the two sets of Darboux coordinates $(u_j,\pi_j)$ and $(\bar u_j,\bar \pi_j)$.
    \item We use this embedding in the perspective neutral structure to find the expression of one set of coordinates in terms of the other, i.e.
    
    \begin{equation}
    \begin{aligned}
    \bar u_i &= a_{ij}u_j\\ 
    \bar \pi_i &= b_{ij}\pi_j.
    \end{aligned}
\end{equation}

    \item To obtain the transformation in the quantum picture we compute the unitary representation of the given canonical transformation, e.g. by the methods derived in \cite{moshinsky1971} or by means of appendix \ref{apsec: the unitary position-to-momentum map}.
\end{enumerate}     
Let us make this procedure explicit in three examples, where we set $m_i=1$ for simplicity.

\subsubsection{Relative position to relative position}
If we want to compute the transformation between two relative position frame, for instance of particle 1 and 2, we write
\begin{align}
\begin{split}
    \begin{pmatrix}
    q_1\\q_2\\q_3
    \end{pmatrix}
    &=
    \begin{pmatrix}
    u_2-u_3\\u_3\\-u_2
    \end{pmatrix}
    =
    \begin{pmatrix}
    -\bar u_3\\\bar u_3-\bar u_1\\\bar u_1
    \end{pmatrix}\\
    \begin{pmatrix}
    p_1\\p_2\\p_3
    \end{pmatrix}
    &= \frac{1}{3}
    \begin{pmatrix}
   \pi_2-\pi_3\\\pi_2+2\pi_3\\-2\pi_2-\pi_3
    \end{pmatrix}
    = \frac{1}{3}
    \begin{pmatrix}
    -\bar \pi_1-2\bar \pi_3\\-\bar \pi_1+\bar \pi_3\\2\bar \pi_1+\bar \pi_3
    \end{pmatrix}.
    \end{split}
\end{align}
From this we can read off the canonical transformation 
\begin{equation}
\label{eq: relative position transformation}
    \begin{aligned}
    u_2 &= -\bar u_1,    &\qquad    u_3 &= \bar u_3-\bar u_1 \\
    \pi_2 &= -(\bar \pi_1+\bar\pi_3) &\qquad  \pi_3 &= \bar\pi_3.  
    \end{aligned}
\end{equation}

Giacomini et.al \cite{giacomini2019} found that the transformation \ref{eq: relative position transformation} can be represented by the unitary operator 
\begin{equation}
    \begin{aligned}
    \label{eq: unitary relative position transformation}
    &H_2 \otimes H_3 \rightarrow H_3 \otimes H_1\\
    &\hat S_{q1\rightarrow q2} = \hat{\mathcal{Q}}_{2,1}e^{i\hat u_2 \hat \pi_3}
    \end{aligned}
\end{equation}
where $\hat{\mathcal{Q}}_{A,B}$ is the position parity-swap operator, acting on vectors in $H_A \otimes H_B$ as
\begin{equation}
    \hat{\mathcal{Q}}_{A,B} \ket{x}_A\otimes\ket{\Psi}_B = \ket{\Psi}_B\otimes\ket{-x}_C.
\end{equation}

\subsubsection{Relative momentum to relative momentum}
We write out the relative momentum representations of particle 1 and 2:
\begin{align}
\begin{split}
            \begin{pmatrix}
                q_1\\q_2\\q_3
            \end{pmatrix}
            &=\frac{1}{3}
            \begin{pmatrix}
                u_2-u_3\\u_2+2u_3\\-2u_2-u_3
            \end{pmatrix}
            =\frac{1}{3}
            \begin{pmatrix}
             -\bar u_1-2\bar u_3\\-\bar u_1+\bar u_3\\2\bar u_1+\bar u_3
            \end{pmatrix}\\
            \begin{pmatrix}
                p_1\\
                p_2\\
                p_3
            \end{pmatrix}
            &= 
            \begin{pmatrix}
                \pi_2-\pi_3 \\
                \pi_3\\
                -\pi_2
            \end{pmatrix}
            =
            \begin{pmatrix}
                \bar\pi_3 \\
                \bar\pi_3-\bar\pi_1\\
                \bar\pi_1
            \end{pmatrix}.
\end{split}            
\end{align}
Again, we read off the canonical transformation
\begin{equation}
\label{eq: relative momentum transformation}
    \begin{aligned}
     u_2 &= -(\bar u_1+\bar u_3)   &\qquad    u_3 &= \bar u_3\\
    \pi_2 &= - \bar \pi_1,    &\qquad    \pi_3 &= \bar \pi_3-\bar \pi_1.  
    \end{aligned}
\end{equation}
and the associated unitary
\begin{equation}
\label{eq: unitary relative momentum transformation}
    \begin{aligned}
    &H_2 \otimes H_3 \rightarrow H_3 \otimes H_1\\
    &\hat S_{p1\rightarrow p2} = \hat{\mathcal{P}}_{2,1}e^{i\hat \pi_2 \hat u_3}
    \end{aligned}
\end{equation}
where $\hat{\mathcal{P}}_{A,B}$ is now the momenta parity-swap operator, acting as
\begin{equation}
    \hat{\mathcal{P}}_{A,B} \ket{p}_A\otimes\ket{\Psi}_B = \ket{\Psi}_B\otimes\ket{-p}_C.
\end{equation}

\subsubsection{Relative position to relative momentum}
This time, we compute the transformation between the relative position and relative momentum of particle 1. We write
\begin{align}
\begin{split}
    \begin{pmatrix}
    q_1\\q_2\\q_3
    \end{pmatrix}
    &=
    \begin{pmatrix}
    u_2-u_3\\u_3\\-u_2
    \end{pmatrix}
    =\frac{1}{3}
    \begin{pmatrix}
    \bar u_2-\bar u_3\\\bar u_2+2\bar u_3\\-2\bar u_2-\bar u_3
    \end{pmatrix}
    \\%
    \begin{pmatrix}
    p_1\\p_2\\p_3
    \end{pmatrix}
    &= \frac{1}{3}
    \begin{pmatrix}
   \pi_2-\pi_3\\\pi_2+2\pi_3\\-2\pi_2-\pi_3
    \end{pmatrix}
    = 
    \begin{pmatrix}
    \bar \pi_2-\bar \pi_3 \\
    \bar \pi_3\\
    -\bar \pi_2
    \end{pmatrix}
\end{split}
\end{align}
Again, we read off the canonical transformation
\begin{equation}
\label{eq: relative q to p transformation}
    \begin{aligned}
    u_2 &= \frac{1}{3}(2\bar u_2+\bar u_3)   &\qquad    u_3 &=\frac{1}{3}(\bar u_2 +2\bar u_3)\\
    \pi_2 &= 2\bar \pi_2-\bar \pi_3,    &\qquad     \pi_3 &= -\bar \pi_2+2\bar \pi_3 
    \end{aligned}
\end{equation}
The associated unitary representation is 
\begin{equation}
\label{eq: unitary relative q to p transformation}
\begin{aligned}
    &H_1 \otimes H_2 \rightarrow H_1 \otimes H_2\\
    &\hat S_{q1\rightarrow p1} = e^{-i\frac{\log{3}}{2} (u_2-u_3)(\pi_2-\pi_3)},
\end{aligned}
\end{equation}
see appendix \ref{apsec: the unitary position-to-momentum map} for the detailed derivation. Since we do not change particle, no parity-swap operator appears.

    \section{Towards a first-class theory}
\subsection{Mathematical preliminaries: The abelian conversion method}
In the reduced quantization approach only the intrinsic variables spanning the reduced phase-spaces $\mathcal{P}_{persp}$ are quantized. Thus, to obtain the perspective dependent quantum descriptions and the transformations between them we had to resort to the the classical perspective neutral structure. This is somewhat unfortunate. We would rather have a \emph{quantum perspective neutral structure}, from which we could derive the perspective-dependent descriptions without ever referring to the classical theory. 

Remember, that the non-commutativity of the second-class constraints was the main obstruction to a direct quantization of $\mathcal{P}_{persp}$. If we could somehow turn them first-class, most of our problems would be solved. This is the basic idea behind the \emph{abelian conversion} scheme, first introduced in \cite{batalin1987}. Instead of passing on to the Dirac bracket and reducing the Phase space, we take the opposite direction and extend it even further by introducing new sets of canonical coordinates. On this extended phase-space we then convert the original second class to first class constraints. Canonical quantization of the extended phase is straightforward and will yield the sought-after perspective-neutral structure in the quantum picture. Let us quickly summarise the main ideas.

When converting, we have to keep the physical degrees of freedom constant. But contrary to second class constraints, first class constraints entail a gauge symmetry. Thus, we have to add $2M$ coordinates $\psi_a$ for every $2M$ second class constraint \cite{amorim1996}. We demand that they fulfill the canonical symplectic structure  
\begin{gather}
\label{eq: shorthand canonical relations}
    [\psi_a, \psi_b] = \omega_{ab}, \quad  \omega =
     \begin{pmatrix}
     0&1\\
     -1&0
    \end{pmatrix}\\
    [y_a, \psi_b] = 0 \quad \forall \; a,b,
\end{gather}
where we have condensed the old positions and momenta into a single variable, writing 
$$
q_i = y_n, \;n = 1\;...\,N, \qquad p_i = y_n, \;n = N+1\;...\;2N.
$$ 
     
On the extended phase space $\mathcal{P}_{ext}$, coordinatized by $(y_n, \psi_m)$, the original second class constraints are converted into abelian\footnote{The condition that the new constraints need to have an abelian structure is not mandatory. In the non-abelian case, equation \ref{eq: abelianized constraints} can have the more general form $[\Tilde{\Phi}_\alpha,\Tilde{\Phi}_\beta] = c_{\alpha, \beta}^\gamma \Tilde{\Phi}_\gamma$.} first class constraints $$\Tilde{\Phi}_\alpha = \Tilde{\Phi}_\alpha(y, \psi)$$ by requiring that they fulfill the differential equations
\begin{equation}
\label{eq: abelianized constraints}
    [\Tilde{\Phi}_\alpha,\Tilde{\Phi}_\beta] = 0\quad \forall \; \alpha,\beta
\end{equation}
with the initial conditions
\begin{equation}
\label{eq: abelianization initial conditions}
    \Tilde{\Phi}_\alpha(y, \psi = 0) = \Phi(y).
\end{equation}

We must further ensure that the extended gauge-system is dynamically equivalent to the original second-class system. This is achieved by a specific extension of the dynamical phase-space functions $F(y) \rightarrow \tilde F(y, \psi)$ \cite{amorim1996}. The abelianized functions $\tilde F(y, \psi)$ have to be gauge invariant with respect to all converted constraints 
\begin{equation}
\label{eq: abelianized functions}
    [\tilde \Phi_\alpha, \tilde{F}] = 0
\end{equation}
and for $\psi =0$ have to coincide with the old ones 
\begin{equation}
\label{eq: abelianized functions initial conditions}
    \tilde F(y, \psi = 0) = F(y).
\end{equation}
The initial conditions \ref{eq: abelianization initial conditions} and \ref{eq: abelianized functions initial conditions} ensure that by gauge fixing the system with $\psi_a= 0$ we regain the equations of motion of the original theory. 

Finding solutions to the equations \ref{eq: abelianized constraints} - \ref{eq: abelianized functions initial conditions} turns out to be highly nontrivial. Fortunately, we do not need to discuss the general case. Amorim and Das showed in \cite{amorim1994} that when the original constraints are linear in the phase space coordinates the abelianized constraints can be written as 
\begin{equation}
\label{eq: abelianized constraints linear}
    \tilde \Phi_\alpha = \Phi_\alpha + X_{\alpha a}\psi^a,
\end{equation}
where
\begin{equation}
\label{eq: condition on X}
    X_{\alpha a}X_{\beta b} \omega^{ab} = -C_{\alpha \beta}.
\end{equation}

In this case, the computation of the converted phase-space functions is much simpler as well. We find that a set of gauge-independent Dirac observables is given by 
\begin{equation}
\label{eq: converted coordinates}
   \tilde y^{c} = y^c-\psi^b \omega _{ab}X^{\alpha a}[\Phi_\alpha, y^c]_y = y^c- \psi^b B^c_b.
\end{equation}
Due to the linearity of the coordinate functions, the matrix $B$ is constant. The $\tilde y^c$ commute by construction with all abelianized constraints
\begin{align}
\begin{split}
[\tilde \Phi_\alpha, \tilde y^c] =&\; [\Phi_\alpha, y^c]  - [\Phi_\alpha,\psi^b]B^c_b 
-X_{\alpha a}B^c_b[\psi^a,\psi^b] \\
=&\; [\Phi_\alpha, y^c] - X_{\alpha a}B^c_b\omega^{ab}\\
=& \;0
\end{split}
\end{align}
Any observable, e.g. the Hamiltonian, defined by the replacement 
$$
\tilde H(y,\psi) = H(\tilde y)
$$
inherits this property and becomes also first class. The total Hamiltonian is obtained by adding the first-class constraints to $\tilde H$, yielding 

$$
\tilde H_T(y,\psi) = H(\tilde y) + \sum_i \lambda_i\tilde\Phi_i.
$$

\subsection{The extended phase-space}
Let us employ the Abelianization scheme for our system at hand. We start with relation \ref{eq: condition on X}, which in matrix form can be written as
\begin{equation}
X\omega X^T = -C
\end{equation}
Carrying out the matrix multiplication we obtain the following equations for the elements of X
\begin{align}
\begin{split}
    \begin{pmatrix}
    a&b\\
    c&d
    \end{pmatrix}
    &\begin{pmatrix}
    0&1\\
    -1&0
    \end{pmatrix}
    \begin{pmatrix}
    a&c\\
    b&d
    \end{pmatrix}
    \\&=
    \begin{pmatrix}
    0&-cb+da\\
    -ad+cb&0
    \end{pmatrix}
    =
    \begin{pmatrix}
    0&-\mu\\
    \mu&0
    \end{pmatrix}.
    \end{split}
\end{align}
So the general form of $X$ is given by
\begin{equation}
    X =
    \begin{pmatrix}
    a&b\\
    c&d
    \end{pmatrix}
    \quad \text{where} \quad cb-da = \mu
\end{equation}
This is just a rescaling of the defining condition $da -cb = 1$ for 2D canonical transformations \cite{moshinsky1971}. We denote the new set of canonical variables extending the original phase-space by $(p,q)$. With \ref{eq: abelianized constraints linear} we obtain the transformed constraints
\begin{align}
\label{eq: transformed constraints}
    \tilde \Phi_1 &= \sum_i q_i +aq+bp\\
    \tilde \Phi_2 &= \sum_i \frac{p_i}{\mu_i}+ cq+dp.
\end{align}
acting on the extended 8-dimensional phase space. They are proofed to be first class when the condition $cb-da = \mu$ holds. Now we compute the functions $\tilde y$ via the equation \ref{eq: converted coordinates}. $B$ is given in matrix form by
\begin{widetext}
\begin{align}
B&= \omega^{-1} X^{-1} \norm{[\Phi_\alpha, y^c]}\\
 &= 
    \begin{pmatrix}
    0&-1\\
    1&0
    \end{pmatrix}
    \frac{1}{-\mu}
     \begin{pmatrix}
    d&-b\\
    -c&a
    \end{pmatrix}
     \begin{pmatrix}
    0&0&0&1&1&1\\
    -{\mu_1}^{-1}&-{\mu_2}^{-1}&-{\mu_3}^{-1}&0&0&0
    \end{pmatrix}\\[4pt]
    &=
    -\frac{1}{\mu}
    \begin{pmatrix}
    a\,{\mu_1}^{-1}&a\,{\mu_2}^{-1}&a\,{\mu_3}^{-1}&c&c&c\\[4pt]
    b\,{\mu_1}^{-1}&b\,{\mu_2}^{-1}&b\,{\mu_3}^{-1}&d&d&d.
    \end{pmatrix}
\end{align}
\end{widetext}
where $\norm{a_{ij}}$ denotes the matrix with components $a_{ij}$. We can thus compute the Dirac observables
\begin{align}
\tilde q_i &= q_i + \frac{1}{\mu\mu_i} (aq+bp)& i &= 1,2,3
\end{align}
and
\begin{align}
\tilde p_i &= p_i + \frac{1}{\mu} (cq+dp)& i &= 1,2,3.
\end{align}
The Hamiltonian becomes
\begin{equation}
    \tilde H = \frac{1}{2}\sum_i \frac{(p_i + \nicefrac{(cq+dp)}{\mu})^2}{\mu_i} + \lambda_1\tilde \Phi_1 + \lambda_2\tilde \Phi_2.
\end{equation}

This is the most general form of the extended structure. However, the explicit representation of the extended phase space does not matter, as $[aq+bp,cq+dp]=ad-bc=-\mu$. So before we continue, let us choose $a=1, d = -\mu, b=c=0$ to simplify the notation. With this, the extended constraints read 
\begin{align}
\label{eq: extended system}
    \tilde{\Phi}_1 &= \sum_i {q}_i + q \simeq 0\\
    \tilde{\Phi}_2 &= \sum_i \frac{{p}_i}{\mu_i}-\mu p = \sum_i \frac{{p}_i- p}{\mu_i} \simeq 0, 
\end{align}
the Dirac observables 
\begin{align}
    \tilde q_i &= q_i + \frac{ q}{\mu\mu_i}\\
    \tilde p_i &= p_i - p
\end{align}
and the Hamiltonian
\begin{align}
    \label{eq: converted hamiltonian}
    \tilde H &= \frac{1}{2}\sum_i \frac{(p_i  -p)^2}{\mu_i} + \lambda_1\tilde \Phi_1 + \lambda_2\tilde \Phi_2.
\end{align}

\subsection{Halfway down: The intermediate phase-spaces}
The 8 dimensional extended phase space $\mathcal{P}_{ext}$ was explicitly constructed so that the original  phase space is reestablished by gauge fixing $q\simeq p \simeq 0$. There is no need, however, to do this simultaneously. We can fix either one of the gauges and define a Dirac bracket on the such obtained 6-dimensional second-class constraint surface. Passing on to intrinsic coordinates then yields an symplectomorphic set of intermediate phase spaces $\mathcal{P}_{int}$ which still feature one first-class constraint.

Let us at first partially gauge fix the extended system by $\Phi_3 =p\simeq0$. We now have a mixed theory featuring one first-class constraint $\Phi_2$ and two second class constraints $\Phi_1$ \& $\Phi_3$. With $[\Phi_1,\Phi_3] = 1$ the Dirac-bracket reads
\begin{equation}
\begin{aligned}
    \{q_i,p_j\}&=\delta_{ij}&\qquad\{q,p\}&=0\\
    \{q_i,p\}&=0&\qquad\{q,p_j\}&=0.\\
\end{aligned}
\end{equation}
Again, we can go over to intrinsic Darboux coordinates defined by $q= -\Phi_1$ and $p = 0$. Both $q$ and $p$ drop out and we are left with the \emph{intermediate momentum space} $\mathcal{P}_{int,p}$, characterised by
\begin{equation}
\begin{aligned}
\label{eq: intermediate momentum space}
     \tilde q_{p,i} &= q_i - \frac{\Phi_1}{\mu\mu_i}&\qquad \Phi_2 &= \sum \frac{p_i}{\mu_i} \simeq 0 \\
     \tilde p_{p,i}  &= p_i&\qquad H_{p} &= H(p_i)+\lambda'_2\Phi_2.
\end{aligned}
\end{equation}
This structure is symplectomorphic to the classical perspective neutral structure established in \cite{vanrietvelde2020}, connected via the canonical transformation \ref{eq: transformation vanrietvelde viktor}. We could now fix the gauge with $q_i \simeq 0$ to obtain internal perspectives symplectomorphic to those found in \cite{vanrietvelde2020}. However, only fixing the gauge with $\Phi_1 \simeq 0$ reinstates the relative meaning of the coordinates and brings us back to $\mathcal{P}_{orig}$. We see that the gauge-freedom exploited in \cite{vanrietvelde2020} can be replaced by a preferred gauge and the freedom to choose internal Darboux-coordinates.

If we instead gauge fix by $q\simeq 0$ first, we obtain the \emph{intermediate position space} $\mathcal{P}_{int,q}$ 
\begin{equation}
\begin{aligned}
\label{eq: intermediate position space}
     \tilde q_{q,i} &= q_i &\quad\Phi_1 &= \sum {q}_i \simeq 0\\
     \tilde p_{q,i} &= p_i - \frac{\Phi_2}{\mu} &\quad H_{q} &=  H(p_i)-\frac{1}{2\mu}\Phi_2^2 + \lambda_1'\Phi_1.
\end{aligned}
\end{equation}
It is easy to see that gauge-fixing the system a second time by $\Phi_2 \simeq 0$ leads us back to original phase space.

It remains to check if the spaces \ref{eq: intermediate position space} and \ref{eq: intermediate momentum space} are symplectomorphic. Indeed, we find that they are connected by the transformation
\begin{equation}
\label{eq: unitary q to p transformation}
    \begin{aligned}
    S_{q\rightarrow p}:\quad &\mathcal{P}_{int,q}\rightarrow \mathcal{P}_{int,q}\\
    &p_i - \frac{1}{\mu}\sum \frac{p_j}{\mu_j} \rightarrow \bar p_i\\
    &q_i \rightarrow \bar q_i -\frac{1}{\mu\mu_i}\sum \bar q_j,
    \end{aligned}
\end{equation}
which can be shown to be canonical by
\begin{align}
\begin{split}
    \ptd[\bar p_i]{p_j} &= \delta_{ij} - \frac{1}{\mu\mu_j} = \ptd[q_i]{\bar q_j}\\
    \ptd[\bar p_i]{q_j} &= \delta_{ij} - \frac{1}{\mu\mu_j} = \ptd[p_i]{\bar q_j}.
\end{split}
\end{align}

    \section{Dirac Quantization of first class constraints}
\label{chap: Quantization of first class constraints}
\subsection{Mathematical preliminaries}

Having worked through the Abelianization scheme we have obtained a pure first-class Hamiltonian system. Various methods of quantizing such a system are known (see \cite{henneaux1994}, chapter 13, for an overview).
An alternative can be found in the \emph{Dirac quantization}. Here one quantizes the full extended phase-space including redundant degrees of freedom and solves the constraints in the quantum theory. To this end we promote all coordinates on $\mathcal{P}_{ext}$ to operators acting on an \emph{extended Hilbert Space} $\H_{ext}$ and require that the physical states are zero eigenstates of the constraints
\begin{equation}
     \label{eq: quantum constraint condition}
         \hat \Phi_\alpha\ket{\phi}_{phys} = 0.
     \end{equation}
This condition is tantamount to requiring that physical states are invariant under gauge transformations 
\begin{equation}
\label{eq: gauge invariance}
    e^{i\epsilon_\alpha \hat \Phi_\alpha} \ket{\phi}_{phys} =  \ket{\phi}_{phys}\quad \forall\alpha\;\; \text{(no sum)}
\end{equation}
generated by the constraints \cite{henneaux1994}. We can find solutions by \emph{group averaging} \cite{marolf2000} via the projector
\begin{align}
\begin{split}
\label{eq: group averaging}
    \delta_\Phi: \frac{1}{4\pi^2}\underbrace{\int_{-\infty}^{+\infty} e^{is_1\Phi_1}ds}_{= 2\pi\delta_{\Phi_1}}\underbrace{\int_{-\infty}^{+\infty} e^{it\Phi_2}dt}_{= 2\pi\delta_{\Phi_2}}\ket{\chi}_{ext}= \ket{\chi}_{phys},
\end{split}
\end{align}
where we used that when $[\hat A,\hat B] = 0$ the identity $e^{\hat A\hat B} = e^{\hat A} e^{\hat B}$ holds such that we can project on each constraint individually by $\delta_{\Phi_i}$.

This projective method is also used in quantum information where it is usually referred to as "G-twirling \cite{palmer2014}. In QI one deals mostly with compact groups, in which case the projector is well defined. However, the group associated with the constraint \ref{eq: quantum constraint condition} is non-compact and the (improper) projector \ref{eq: group averaging} does not converge on $\H_{ext}$.  In other words, the physical states are not actually contained in $L^2(\R^4)$ as they are not square-integrable. A "home" for the projector and the physical state can be found in the so-called rigged Hilbert space \cite{delamadrid2005, delamadridmodino2001, thiemann}, with the hermitian and (presumably) positive definite scalar product \cite{marolf2000}
\begin{align}
\begin{split}
\left(\delta_\Phi f, \delta_\Phi g  \right)_{phys} &= [\delta_\Phi(g)](f)\widehat{=} \bra{f}\delta_\Phi\ket{g}\\ 
\forall f,&g\in\mathcal{D}_{ext}.
\end{split}
\end{align}

\subsection{The perspective-neutral structure in the quantum picture}
Putting hats on the observables, equations \ref{eq: extended system} - \ref{eq: converted hamiltonian} comprise what we call the quantum perspective-neutral structure. A quantum state in the extended Hilbert space $\H_{ext}$, for example given in momentum representation by 
\begin{equation}
    \ket{\chi}_{ext} = \int \chi(\vec{p},p)\ket{\vec{p}}\ket{p}d\Vec{p}dp.
\end{equation} 
has 4 degrees of freedom, twice as many as physically observable. Again, these redundancies must be fixed to re-obtain the perspective-dependent descriptions. Here we follow largely the quantum symmetry reduction scheme of Vanrietvelde et al.: We define a unitary trivialization map $\hat T$ that transforms one constraint in such a way that it acts only on one variable.\footnote{In the literature, such trivialization maps were employed in different contexts. E.g., see \cite{creutz_muzinich_tudron_1979} for their use in the canonical quantization of non-Abelian gauge fields and \cite{anderson_1993} for an appliance related to quantum canonical transformations.} After an (improper) projection on this trivialized constraint we can discard the now redundant degrees of freedom. As we are dealing we two constraints we need two trivialization maps, each followed by a projection. 

It seems sensible to get rid of the additional 4th degree of freedom at the very beginning, as how to physically interpret it remains - at least to the author - rather obscure. To this end, we can choose to project on any linear combination of the constraints, given that it is rotated to act only one the $(q,p)$-slot. We thus obtain one of many \emph{intermediate Hilbert spaces} $\H_{int}$, which are the quantum analogues of the intermediate phase spaces. By repeating the procedure - trivialization followed by projecting - for a coordinate of our liking we arrive at the perspective dependent description. We will now have a look at the symmetry reduction scheme in detail.

\subsection{Relative position and momentum frame revisited}
\subsubsection{The intermediate and relative position frames}
Let us start with an initial projection on $\hat p = 0$. We therefore need to define the trivialization map
\begin{equation}
    \that_p= e^{-\frac{i}{\mu}\hat q\hat \Phi_2}
\end{equation}
which rotates the constraints to 
\begin{align}
    \label{eq: constraints 1st projection}
    \that_p \hat{\tilde{\Phi}}_1 \thatdagger_p &= \hat \Phi_1 = 0,&
    \that_p \hat{\tilde{\Phi}}_2 \thatdagger_p &= \hat p = 0.
\end{align}
Acting with $\that_p$ on a general state in $\H_{ext}$ following up with the projector $\delta_{T {\tilde{\Phi}}_2 T^\dagger} = \delta_p$ yields
\begin{multline}
\label{eq: first projection}
   \delta_p \that_p \ket{\chi}_{ext} =\\ 
   \int \chi(\vec{p},\sum_i\frac{p_i}{\mu_i\mu})\ket{p_1}\ket{p_2}\ket{p_3}d\Vec{p}\otimes \ket{p=0}.
\end{multline}

We can use this result to compute the action of $\delta_p\that_p$ on the Dirac observables and on the Hamiltonian \ref{eq: converted hamiltonian}. From $\hat p\, \delta_p \that_p \ket{\chi}_{ext}=0$ it follows that 
\begin{gather}
\begin{split}
    _{ext}\bra{\chi}(T_p)^\dagger(\delta_p)^\dagger \hat p\, \delta_p \that_p \ket{\chi}_{ext}=0\\
    \rightarro (\delta_p)^\dagger \hat p \delta_p = 0
    \end{split}
\end{gather}
So after acting with $T_p$ we can discard all terms containing $\hat p$, which leaves us with
\begin{align}
\begin{split}
    \hat{\tilde q}_i & \rightarro \hat q_{q,i} = \hat q_i \\
    \hat{\tilde p}_i & \rightarro \hat p_{q,i} = \hat p_i -\frac{\hat\Phi_2}{\mu} \\
    \hat{\tilde {H}} & \rightarro \hat H_{q} = \hat{H}(p_{q,i})\\ 
    &\hspace{4.9em}= \hat H(p_i)-\frac{1}{2\mu}\hat\Phi_2^2 + \hat\lambda_q \hat\Phi_1
    \label{eq: quantum intermediate position Hamiltonian}
    \end{split}
\end{align}

Now all information is stored in the first three slots and as in the classical case $\hat q$ and $\hat p$ have dropped entirely from the formalism. It is therefore permissible to discard the 4th degree of freedom altogether by projecting onto the classical gauge fixing condition
\begin{equation}
\begin{aligned}
    \ket{\chi}_{q} &=\sqrt{2\pi} \bra{q = 0} \delta_p \that_p \ket{\chi}_{ext}=                       \int dp \braket{p|\,\delta_p \that_q|\chi}_{ext}\\
                     &= \int \chi_{q}(\vec{p})\ket{p_1}\ket{p_2}\ket{p_3}d\Vec{p}.
\end{aligned}
\end{equation}
where  
$$\chi_{q}(\vec{p})= \chi(\vec{p}, \sum_i \frac{p_i}{\mu_i\mu}).$$ 
The structure above is the quantum version of \ref{eq: intermediate position space}. We therefore speak of the \emph{intermediate position space} $\H_{int,q}$. 

To dispose of the remaining redundancy and obtain the perspective-dependent quantum state we must carry out the symmetry reduction once more. Contrary to the first projection, we are now free to choose the slot we want to get rid of. This amounts to defining another trivialization map, such that the transformed constraint acts only on the d.o.f. in question. For example, to get rid of the first slot we apply
\begin{equation}
   \hat{T}_{q1} = e^{-i\frac{\pi}{2}(\hat q_2 \hat p_3 - \hat q_3 \hat p_2)} e^{-i\hat p_1(\hat q_2+\hat q_3)}
\end{equation}
While the second exponential is the actual trivialization operator, the first one induces the canonical transformation 
\begin{align}
q_2&\rightarrow q_3 &p_2&\rightarrow p_3\\
q_3 &\rightarrow -q_2&p_3&\rightarrow -p_2. 
\end{align}
This 90°-rotation in the $(q_2, q_3)$- and $(p_2, p_3)$-subspaces is needed to bring the coordinate names in accordance with those used in the reduced quantization scheme. $\hat{T}_{q1}$ rotates the remaining constraint to
\begin{equation}
\hat{T}_{q1}\hat\Phi_1 \hat{T}_{q1}^\dagger = \hat q_1   
\end{equation}

Let us apply $\that_{q1}$ followed by the projection $\sqrt{2\pi} \bra{p_1=0}\delta_{q_1}$ to the Dirac-observables. We thereby obtain a new description in $\H_{persp}$:
\begin{align}
\begin{split}
    \begin{pmatrix}
    q_1\\
    q_2\\
    q_3
    \end{pmatrix}
    &\rightarrow
    \begin{pmatrix}
    q_2-q_3\\
    q_3\\
    -q_2
    \end{pmatrix}\\
    \begin{pmatrix}
    p_{q,1}\\
    p_{q,2}\\
    p_{q,3}
    \end{pmatrix}
    &\rightarrow \frac{1}{\mu}
    \begin{pmatrix}
    \frac{p_2}{\mu_3}-\frac{p_3}{\mu_2}\\
    \frac{p_2}{\mu_3}+ p_3(\mu -\frac{1}{\mu_2})\\
    -p_2(\mu -\frac{1}{\mu_3})-\frac{p_3}{\mu_2}
    \end{pmatrix}.
    \end{split}
\end{align}
Plugging this into the Hamiltonian \ref{eq: quantum intermediate position Hamiltonian} we arrive at
\begin{multline}
   \hat{H}_q  \rightarrow  \hat H_{q1} = \frac{1}{2\mu}\Big(
    (\frac{\mu}{\mu_3}-\frac{1}{\mu_3^2})\hat p_2^2 +
    (\frac{\mu}{\mu_2}-\frac{1}{\mu_2^2})\hat p_3^2 + 2\frac{\hat p_2\hat p_3}{\mu_2\mu_3}\Big).
\end{multline}

This is equal to the relative position Hamiltonian \ref{eq: relative position hamiltonian} from the perspective of particle 1, which we obtained through the reduced-quantization method. There are no constraints or redundancies left. We can summarise the whole process by
\begin{equation}
    \ket{\chi}_{qi} = \bra{p_i=0}\bra{q=0}\delta_{q_i}\delta_p T_{q_i}T_p\ket{\chi}_{ext}. 
\end{equation}

\subsubsection{The intermediate and relative momentum frames}
Now let us project on  $\hat q = 0$ first. In this case we apply the trivialization map
\begin{equation}
    \that _q= e^{-i\hat p\hat \Phi_1}
\end{equation}
which rotates the constraints to
\begin{align}
    \label{eq: constraints 1st projection momentum}
    \that _q \hat{\tilde{\Phi}}_1 \that _q^\dagger &= \hat q=0&
    \that _q \hat{\tilde{\Phi}}_2 \that _q^\dagger &= \hat \Phi_2=0.
\end{align}
This rotation, followed by the projection $\sqrt{2\pi} \bra{p=0}\delta_{\hat q}$ transforms the Dirac-coordinates and the Hamiltonian to
\begin{align}
\begin{split}
    \hat{\tilde q}_i & \rightarro \hat q_{p,i} = \hat q_i - \frac{\Phi_1}{\mu\mu_i} \\
    \hat{\tilde p}_i & \rightarro \hat p_{p,i} = \hat p_i \\
    \hat{\tilde {H}} & \rightarro \hat H_p = \hat{H}(p_{p,i}) = \hat H+\lambda'_2\Phi_2.
\end{split}
\end{align}

It comes as no surpise, that the above structure is the quantum version of the classical intermediate momentum space. To obtain the perspective-dependent description we must rotate the system once more, this time by
\begin{equation}
   \that _{p1}= e^{-i\frac{\pi}{2}(\hat q_2 \hat p_3 - \hat q_3 \hat p_2)} e^{i\mu_1q_1(\frac{\hat p_2}{\mu_2}+\frac{\hat p_3}{\mu_3})}
\end{equation}
and then project by $\sqrt{2\pi} \bra{p_1=0}\delta_{p_1}$. This transforms the Dirac-coordinates to
\begin{align}
\begin{split}
    \begin{pmatrix}
    q_{p,1}\\
    q_{p,2}\\
    q_{p,3}
    \end{pmatrix}
    &\rightarrow \frac{1}{\mu}
    \begin{pmatrix}
    \frac{1}{\mu_1}(q_3-q_2)\\
    \frac{q_2}{\mu_3}+ q_3(\mu -\frac{1}{\mu_2})\\
    -q_2(\mu -\frac{1}{\mu_3})-\frac{q_3}{\mu_2}.
    \end{pmatrix}\\
    \begin{pmatrix}
    p_1\\
    p_2\\
    p_3
    \end{pmatrix}
    &\rightarrow
    \begin{pmatrix}
    \mu_1(\frac{p_2}{\mu_3}-\frac{p_3}{\mu_2})\\
    p_3\\
    -p_2.
    \end{pmatrix}.
\end{split}    
\end{align}
The Hamiltonian becomes
\begin{multline}
    \hat{H}_p  \rightarrow  \hat{H}_{p1} = \frac{1}{2}\Big(\frac{\mu_1+\mu_2}{\mu_2^2}\hat p_3^2\\ + \frac{\mu_1+\mu_3}{\mu_3^2}\hat p_2^2-\frac{2\mu_1}{\mu_2\mu_3}\hat p_2\hat p_3\Big).
\end{multline}

We have therefore arrived at the relative-momentum description from the perspective of particle one. We can again condense the process to

\begin{equation}
    \ket{\chi}_{pj} = \bra{q_j=0}\bra{p=0}\delta_{p_j}\delta_q T_{p_j}T_q\ket{\chi}_{ext}. 
\end{equation}

 \subsection{Changing quantum reference frames via the perspective neutral structure}
Now let us see how to change reference frames via the the quantum perspective-neutral theory. We will focus on the three cornerstone-transformation of type \ref{eq: unitary relative position transformation}, \ref{eq: unitary relative momentum transformation} and \ref{eq: unitary relative q to p transformation}, by whose combination one can compose arbitrary maps. To transform between relative position frames we make use of the intermediate position space. That means, to recover transformations like \ref{eq: unitary relative position transformation}  we at first invert the quantum symmetry reduction and re-embed $\H_{q1}$ into $\H_{q}$, followed by a projection on $\H_{q2}$. Concretely, in \cite{vanrietvelde2020}, it is proofed that this map is given by
\begin{equation}
    \hat S_{qi\rightarrow qj} := \int dq_j \bra{q_j} \hat T_{qj} \,  (\hat T_{qi})^\dagger \ket{q_i = 0} \otimes [\cdot]
\end{equation}
where the reduced state $\ket{\chi}_{q1}$ must be inserted into the empty slot $[\cdot]$. 

Likewise, a transformation between two relative momentum maps can be implemented by
\begin{equation}
    \hat S_{pi\rightarrow pj} := \int dp_j \bra{p_j} \hat T_{pj} \,  (\hat T_{pi})^\dagger \ket{p_i = 0} \otimes [\cdot].
\end{equation}

Observe, that via the intermediate spaces we can switch between particle-perspectives but not between relative position and momentum representation. To make this possible we have to go all the way up and make use of the extended Hilbert space.  Transformations of the type \ref{eq: unitary relative q to p transformation} are thus implemented by
\begin{widetext}
\begin{equation}
    \hat S_{qi\rightarrow pi} := \int dq dp_j \bra{p_j} \hat T_{pj} \bra{q=0}\hat T_q(\hat T_p)^\dagger \ket{p=0}\otimes(\hat T_{pi})^\dagger \ket{q_i = 0} \otimes [\cdot].
\end{equation}
\end{widetext}
    \label{chap: conclusio}
\section{Recap and Comparison}
Let us review our framework's mathematical structure and and compare it to that of Vanrietvelde et al.. Both frameworks originated from the same translational invariant, toy model Lagrangian. The re-derivation of some of the results was, therefore, expected and desired.  While in \cite{vanrietvelde2020} the Legendre transformation was performed on the original, gauge dependent variables, in this article we first introduced gauge-invariant, relative variables. This mapping entailed a constraint of the relative positions. The Legendre transformation led to the known constraint for the sum of relative momenta, yielding a symmetrical Hamiltonian structure living in $\mathcal{P}_{orig}$. 

As the constraints were second-class, there was no gauge freedom to exploit and we could not obtain the perspective dependent descriptions by gauge fixing. Instead we showed that the perspective dependent descriptions can be understood as Darboux coordinates on the constraint surface. This approach allowed us to derive a broader class of perspective dependent frames. In addition to the already known relative momentum frames $\mathcal{P}_{pi}$ we derived the new relative position frames $\mathcal{P}_{qj}$. 
\begin{figure*}
\hspace*{-0.5cm} 
\centering
\begin{tikzpicture}

        \node (ext) [outer sep=5pt] at (-3,10) {$\mathcal{P}_{ext} \sim \R^8$};
        \node (auxq) [outer sep=5pt] at (-4,7) {$\mathcal{P}_{int, q}$};
        \node (auxp) [outer sep=5pt] at (-2,7) {\textcolor{orange}{$\mathcal{P}_{int, p}$}};
        \node [outer sep=5pt] at (-3,7) {$\sim$};
        \node (orig) [draw, outer sep=5pt] at (-3,4) {$\mathcal{P}_{orig} \sim \R^6$};
        \node (persp) [outer sep=5pt] at (-3,1) {$\mathcal{P}_{qi} \sim \R^4\sim$\textcolor{orange}{$\; \mathcal{P}_{pi}$}};
        \node (dic) [align=right] at (-5.5,2.5) {Darboux\\ coordinates};
        
        \draw  [->] (ext) to node [left] {$q=0$}(auxq) ;
        \draw  [->] (ext) to node [right] {$p=0$}(auxp);
        \draw  [->] (auxq) to node [left] {$\Phi_2\simeq0$}(orig);
        \draw  [->] (auxp) to node [right] {$\Phi_1\simeq0$}(orig);
        \draw  [->] (orig.west) to  [bend left = 40] node [left, rotate=90, anchor = south]{Abelianization}(ext.west);
        \draw  [->] (orig) to node [left] {} ([xshift=-0.8 cm]persp.north);
        \draw  [->] (orig) to node [left]{} ([xshift=0.8 cm]persp.north);
        \draw  [orange,->] (auxp.south east) to  [bend left = 40] node [right, rotate=270, anchor = south]{\textcolor{orange}{$q_i=0$}}(persp.north east);

\node (hext)  [outer sep=5pt] at (5,10)  {$\mathcal{H}_{ext}$};
\node (hauxq) [outer sep=5pt] at (6,5.5) {$\mathcal{H}_{int,q}$};
\node (hauxp) [outer sep=5pt] at (4,5.5) {\textcolor{orange}{$\mathcal{H}_{int,p}$}};
\node [outer sep=5pt] at (5,5.5) {$\sim$};
\node (hpersp)[outer sep=5pt] at (5,1)   {$\textcolor{orange}{\mathcal{H}_{pi}} \sim \H_{qi}$};
\draw  [->] (hext) to node [right] {$\bra{q=0}\delta_p T_p$}(hauxq) ;
\draw  [->] (hext) to node [left] {$\bra{p=0}\delta_q T_q$}(hauxp);
\draw  [->] (hauxq) to node [right] {$\bra{q_i=0}\delta_{qi} T_{qi}$}([xshift=0.6 cm]hpersp.north);
\draw  [orange,->] (hauxp) to node [left] {\textcolor{orange}{$\bra{p_i=0}\delta_{pi} T_{pi}$}}([xshift=-0.6 cm]hpersp.north);

\draw  [->] (ext) to node [above] {Dirac quantization}(hext);
\draw  [orange,->] (persp) to node [above] {\textcolor{orange}{reduced quantization}}(hpersp);
\draw  [orange,->] (auxp) to (hauxp);
\end{tikzpicture}
    \caption{A summary of all the spaces and their connections making appearance in this article. The orange-coloured parts of the framework have either direct counterparts in  \cite{vanrietvelde2020} ($\mathcal{P}_{pi},\mathcal{H}_{pi}$) or are connected by the symplectomorphism defined in \ref{eq: transformation vanrietvelde viktor} ($\mathcal{P}_{int,p},\mathcal{H}_{int,p}$). This is a visual demonstration of the full integration of the original framework in our broader structure. The horizontal axis indicates the number of unphysical degrees of freedom featured in the theory. The further on top, the more redundant coordinates are present. The strong equations $q,p,q_i =0$ should be understood as including the transition to intrinsic coordinates.}
    \label{fig: full scheme}
\end{figure*}
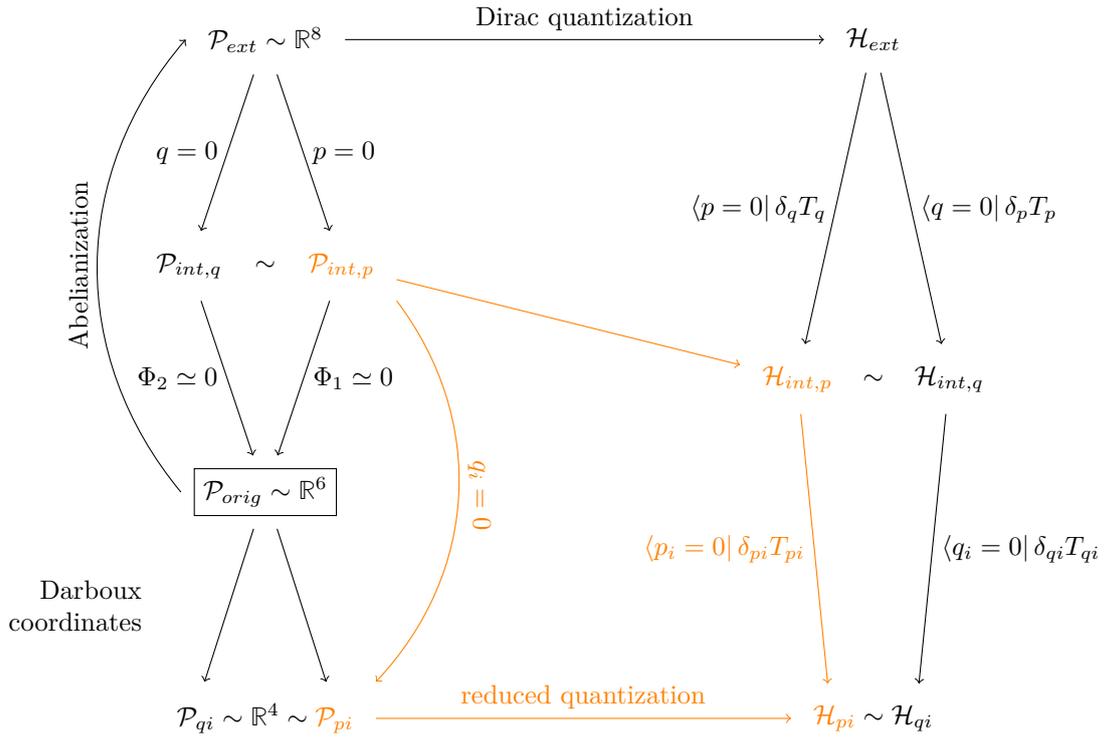
Following this line of thought, we also had to adapt our understanding of reference frame transformations. While in \cite{vanrietvelde2020} reference frame transformations were related to gauge transformations, we showed that they can be understood as Darboux-reparametrizations of the constraint surface. Thus, we were not only able to recover the transformations $\mathcal{P}_{pi}\rightarrow\mathcal{P}_{pj}$ connecting relative momentum frames, but could derive the transformations $\mathcal{P}_{qi}\rightarrow\mathcal{P}_{qj}$ between relative position frames and $\mathcal{P}_{qi}\rightarrow\mathcal{P}_{pi}$ between relative position and relative momentum frames. Canonical quantization of the perspective dependent phase spaces then led us to the perspective dependent Hilbert spaces $\mathcal{H}_{qi}$ and $\mathcal{H}_{pi}$. 

 Our next objective was to find the quantum perspective neutral structure. To this end, we expanded the original phase space to $\mathcal{P}_{ext}$ and modified the constraints to make them first class, a process called Abelianization. Canonical quantization of $\mathcal{P}_{ext}$ yielded the sought-after perspective neutral Hilbert space $\mathcal{H}_{ext}$. To return to the perspective dependent frames, we followed an extended version of the quantum symmetry reduction scheme employed in \cite{vanrietvelde2020}. Having two constraints, two trivialization maps and two projections were needed. Since it was not clear how to interpret the additional degrees of freedom physically, we tried to get rid of them first. We could do this by either projecting on $\hat p=0$ or $\hat q=0$ (or any linear combinations thereof), both options leading to the unitary equivalent, intermediate Hilbert spaces $H_{int, q}$ and $H_{int, p}$. We found that projecting on $\hat q=0$ resulted in a Hilbert space equivalent to the auxiliary Hilbert space of \cite{vanrietvelde2020}. Trivializing and projecting on the remaining constraint then lead us back to the various perspective dependent Hilbert spaces.

We could then make use of the quantum perspective neutral structure   to switch perspectives. Jumping between two relative momentum frames is done by first inverting the second symmetry reduction, thus embedding $\mathcal{H}_{pi}$ into $\mathcal{H}_{p}$ and then projecting onto a different degree of freedom. In an analogous way we could change between two relative position frames. To switch relative position to relative momentum, however, we had to go all the way up and and embed the perspective dependent descriptions in the extended Hilbert space.

till, there is one part missing. Where is the original phase space of \cite{vanrietvelde2020} hidden in our framework? We found that partially  gauge  fixing  $\mathcal{P}_{ext}$ led us to  the intermediate  phase  spaces $P_{int,q}$ and $P_{int,p}$ which could be shown to be isomorphic  to  the  auxiliary  space in \cite{vanrietvelde2020}.  But even though their visual appearance is almost identical, there is an important interpretive difference.  As the coordinates have relative meaning in our framework, fixing the gauge by $q_i= 0$ is physically unjustified.  The only permissible gauge is $\sum q_i\simeq 0$, leading us back to $\mathcal{P}_{orig}$.

\section{Limitations and Outlook}
In the present article, we have further developed the first-principle approach to quantum reference frames introduced by Vanrietvelde et al.. Our construction provides a unifying framework to systematically derive a large class of perspective dependent descriptions and the transformations connecting them. However, it can count only as a very first step towards a  general theory of quantum reference frames.

Some of the necessary generalization are obvious. Our framework covered only a toy model of three free particles in one dimension. As a first step, it should be generalized to multiple particles in 3d space. In \cite{vanrietvelde2018}, it is shown that in this case globally valid gauge conditions are impossible. Since we can regard two second class constraints as mutual gauge fixes, this will likely obstruct the definition of globally valid Darboux coordinates. Further, we should investigate systems with non-vanishing potential energy. It is not obvious that the Dirac bracket will remain its simple form in this case. 

In our simple toy-model, we only obtained a subset of all canonical transformations between quantum reference frames. The limiting factor was the general non-commutativity of canonical transformations and canonical quantization. To not encounter possible mathematical pitfalls, we restricted to linear canonical transformations which do not mix positions and momenta. This excludes the larger part of the extended Galilean transformations, like boosts and transformations to accelerated reference systems, as presented in \cite{giacomini2019}. To extend the framework, a better understanding of the interplay of canonical quantization and canonical transformations is crucial.

Finally, if our aim is to develop a theory about quantum gravity, a generalization to relativistic systems is a plausible next step. Some research papers advance in this direction \cite{giacomini2019a,streiter2021}. But to the author's knowledge, there does not yet exist a comprehensive theory describing transformations between finite-mass reference frames moving with relativistic velocities, not even at the classical level. Only after we have better understood the particularities of such systems from a classical viewpoint we can start working towards a Lorentz-invariant theory of quantum reference frames. 


    \printbibliography
    \onecolumn\newpage
\appendix
\numberwithin{equation}{section}
\renewcommand{\theequation}{\thesection\arabic{equation}}
\section{The Lagrangian in relative coordinates}
\label{apsec:rewriting the lagrangian}
Let us walk through the short computation of the Lagrangian in relative coordinates \ref{eq: Lagrangian new}. We start by expanding the center-of-mass-term and pulling it under the sum, yielding
\begin{equation}
    L(x,\dot{x}) = \frac{1}{2}\sum_{i=1}^3\left( m_i\dot{x}_i^2- \frac{m_i^2}{M}\dot{x}_i^2\right) - \frac{1}{M} \sum_{i < j} m_im_j \dot x_i \dot x_j.
\end{equation}
Let us have a closer look at the first term ($i=1$) of this expression:
\begin{equation}
        m_1\dot{x}_1^2- \frac{m_1^2}{M}\dot{x}_1^2 = \frac{1}{M}\left( m_1(m_1+m_2+m_3) - m_1^2)\right)\dot x_1^2 = \frac{m_1(m_2+m_3)}{M} \dot x_1^2. 
\end{equation}
Expanding both sums we obtain
\begin{multline}
L = \frac{1}{2M}(m_1(m_2+m_3)\dot x_1^2 + m_2 (m_1+m_3)\dot x_2^2 + m_3(m_1+m_2)\dot x_3^2- 2m_1m_2\dot x_1\dot x_2 - 2m_1m_3\dot x_1 \dot x_3 - 2m_2m_3 \dot x_2 \dot x_3 \nonumber
\end{multline}
which we can contract to full squares to arrive at
\begin{equation}
\label{eq: Lagrangian 1}
   L = \frac{1}{2M}\left( m_1m_2(\dot x_1-\dot x_2)^2  + m_1m_3(\dot x_1-\dot x_3)^2 +m_2m_3(\dot x_2-\dot x_3)^2 \right).
\end{equation}
Plugging in the relative coordinates \ref{eq: coordinate transform} and making use of the reduced masses $\mu_i$ then yields the Lagrangian \ref{eq: Lagrangian new}.

\section{Derivation of the consistency conditions}
\label{apsec: derivations of consistency conditions}
This appendix shows the detailed computation of the consistency conditions
\begin{equation}
\label{apeq: consistency}
\dot\Phi_m=[\Phi_m, H_T]\simeq 0   
\end{equation}
and the derivation of the full constraint algebra starting from the Hamiltonian
\begin{equation}
\label{apeq: Hamiltonian}
 H_T = \frac{1}{2}\sum_{i=1}^3 \left(\frac{p_i^2}{\mu_i} - q_4q_i\right) + yp_4  
\end{equation}
and the constraint $\Phi_1 = p_4 \simeq 0$. 

Since $p_4$ commutes with all coordinates but $q_4$, we recover easily constraint \ref{eq: constraint Q} by
\begin{align}
    \dot \Phi_1 = [p_4,H_T] = \sum_{i=1}^3 q_i = \Phi_2 \simeq 0.
\end{align}
Plugging this again in \ref{apeq: consistency} yields
\begin{align}
    \dot\Phi_2 &=  [\sum_{i=1}^3 q_i, \frac{1}{2}\sum_{j=1}^3 \frac{p_j^2}{\mu_j}+ q_4\sum_{i=1}^3q_i + yp_4] \nonumber\\
               &=  \frac{1}{2}\sum_{i,j} \frac{1}{\mu_j}[q_i, p_j^2] 
               = \frac{1}{2}\sum_{i,j}\frac{2}{\mu_j}[q_i, p_j]p_j 
                = \sum_{i,j}\frac{1}{\mu_j}\delta_{ij}p_j\\
               &= \sum_{i=1}^3 \frac{p_j}{\mu_j} = \Phi_3 \simeq 0.
\end{align}
We continue with
\begin{align}
    \dot \Phi_3 &= [\sum_{i=1}^3 \frac{p_j}{\mu_j}, q_4\sum_{i=1}^3q_i]
                = \sum_{i,j} q_4[\frac{p_j}{\mu_j},q_i ] = \sum_{i,j} \frac{q_4}{\mu_j} \delta_{ij}\\
                &= \mu q_4 = \Phi_4 \simeq 0,\\
    \dot \Phi_4 &= [\mu q_4, yp_4] = \mu y[q_4, p_4] \\
                &= \mu y \simeq 0.
\end{align}
Since the last equation gives a condition for the Lagrange-multiplier, the process terminates here. We are left with four constraints
\begin{equation}
 \begin{split}
\Phi_1 &= p_4  \simeq 0,  \\
\Phi_2 &= \sum_iq_i  \simeq 0, \\
\Phi_3 &= \sum_i \frac{p_i}{\mu_i}  \simeq 0,\\
\Phi_4 &= \mu q_4 \simeq 0
\end{split}   
\end{equation}
and one condition for the Lagrange multiplier
\begin{equation}
    y \simeq 0.
\end{equation}

Note that the product of two weak equations $\Phi_i \simeq \Phi_j \simeq 0$ is a strong equation, that is its Poisson bracket
\begin{equation}
    [F,\Phi_i \Phi_j] = [F,\Phi_i] \Phi_j + \Phi_i[F, \Phi_j] \simeq 0
\end{equation}
with an arbitrary function $F(q,p)$ vanishes on the constraint surface. We can therefore discard the last two terms from the Hamiltonian \ref{apeq: Hamiltonian} and simply write
\begin{equation}
    H_T =\frac{1}{2}\sum_i \frac{p_i^2}{\mu_i}.
\end{equation}
Since both $q_4$ and $p_4$ are restricted to zero, they carry no physical information and can be discarded from the formalism \cite{dirac2001}. This means that the system at hand lives in a 6-dimensional phase-space and features two constraints.

\section{The unitary position-to-momentum map}
\label{apsec: the unitary position-to-momentum map}
To find the unitary operator 
\begin{equation}
    \hat S_{q1\rightarrow p1} = e^{\lambda \hat r}
\end{equation}
inducing the canonical transformation
\begin{equation}
\label{apeq: relative position transformation}
    \begin{aligned}
    u_2 &= \frac{1}{3}(2\bar u_2+\bar u_3)   &\qquad    u_3 &=\frac{1}{3}(\bar u_2 +2\bar u_3)\\
    \pi_2 &= 2\bar \pi_2-\bar \pi_3,    &\qquad     \pi_3 &= -\bar \pi_2+2\bar \pi_3
    \end{aligned}
\end{equation}
we first compute the invariants of the transformation
\begin{equation}
    I_1 = u_2 + u_3, \qquad I_2 = \pi_2 + \pi_3.
\end{equation}
In the unitary representation $I_1$ and $I_2$ must also stay invariant \cite{591562}. Using the Hadamard identity we can write
\begin{equation}
    \hat S \hat I_i \hat S^\dagger = \hat I_i + \lambda [\hat r, \hat I_i] + \frac{1}{2!}\lambda^2[\hat r[\hat r, \hat I_i]] + \frac{1}{3!}\lambda^3[\hat r[\hat r[\hat r, \hat I_i]]] + \dots \stackrel{!}{=} \hat I_i.
\end{equation}
We thus see that we must choose $\hat r$ such that the commutator $[\hat r, \hat I_i]$ vanishes. After a little guesswork we find
\begin{equation}
    \hat r = (\hat u_2-\hat u_3)(\hat \pi_2-\hat \pi_3).
\end{equation}
It remains to calculate the phase $\lambda$. To this end we compute e.g.
\begin{equation}
    \hat S \hat \pi_2 \hat S^\dagger = \pi_2 + \frac{\pi_2-\pi_3}{2}(e^{2is}-1).
\end{equation}
Comparing this with \ref{apeq: relative position transformation} we find $s=-i\frac{\log(3)}{2}$. This leaves us with the sought-after unitary operator
\begin{equation}
    \hat S_{q1\rightarrow p1} = e^{-i\frac{\log 3}{2}(\hat u_2-\hat u_3)(\hat \pi_2-\hat \pi_3)}.
\end{equation}
\endgroup

\end{document}